\documentclass{aa}  
\usepackage{xcolor}  

\usepackage{rotating}
\usepackage{graphicx}

\usepackage{lipsum}  

\usepackage{txfonts}
\usepackage{natbib,twoopt}
\usepackage[breaklinks=true]
{hyperref} 
\bibpunct{(}{)}{;}{a}{}{,}             
\makeatletter
  \newcommandtwoopt{\citeads}[3][][]{\href{http://adsabs.harvard.edu/abs/#3}%
    {\def\hyper@linkstart##1##2{}%
     \let\hyper@linkend\@empty\citealp[#1][#2]{#3}}}
  \newcommandtwoopt{\citepads}[3][][]{\href{http://adsabs.harvard.edu/abs/#3}%
    {\def\hyper@linkstart##1##2{}%
     \let\hyper@linkend\@empty\citep[#1][#2]{#3}}}
  \newcommandtwoopt{\citetads}[3][][]{\href{http://adsabs.harvard.edu/abs/#3}%
    {\def\hyper@linkstart##1##2{}%
     \let\hyper@linkend\@empty\citet[#1][#2]{#3}}}
  \newcommandtwoopt{\citeyearads}[3][][]%
    {\href{http://adsabs.harvard.edu/abs/#3}
    {\def\hyper@linkstart##1##2{}%
     \let\hyper@linkend\@empty\citeyear[#1][#2]{#3}}}
\makeatother
\newcommand\gaia {\textit{Gaia}}
\newcommand\kms {$\mathrm{~km~ s^{-1}}$}

\begin{document}

   \title{Searching for compact hierarchical triple system candidates in astrometric binaries and accelerated solutions}

   \subtitle{}
  \titlerunning{Triple Systems in Astrometric Binaries}
  
   \author{D. Bashi
          \inst{1}
          \and
          A. Tokovinin\inst{2}
          }

   \institute{Astrophysics Group, Cavendish Laboratory, University of Cambridge, JJ Thomson Avenue, Cambridge CB3 0HE, UK\\
              \email{db975@cam.ac.uk}
         \and
Cerro Tololo Inter-American Observatory, NSF's NOIRLab, Casilla 603, La Serena, Chile\\
             \email{andrei.tokovinin@noirlab.edu}
             }

   \date{Received XXX. XX, XXXX; accepted YYY. YY, YYYY}

 
  \abstract
   {Compact hierarchical triple (CHT) systems, where a tertiary component orbits an inner binary, provide critical insights into stellar formation and evolution. Despite their importance, the detection of such systems, especially compact ones, remains challenging due to the complexity of their orbital dynamics and the limitations of traditional observational methods.}
   {This study aims to identify new CHT star systems among \gaia \ astrometric binaries and accelerated solutions by analysing the radial velocity (RV) amplitude of these systems, thereby improving our understanding of stellar hierarchies.}
   {We selected a sample of bright astrometric binaries and accelerated solutions from the \gaia\  DR3 Non-Single Stars catalogue. The RV peak-to-peak amplitude was used as an estimator, and we applied a new method to detect potential triple systems by comparing the RV-based semi-amplitude with the astrometric semi-amplitude. We used available binary and triple star catalogues to identify and validate candidates, with a subset confirmed through further examination of the RV and astrometric data.}
   {Our analysis resulted in the discovery of $956$ CHT candidates among the orbital sources as well as another $3,115$ probable close binary sources in stars with accelerated solutions. Exploring the inclination, orbital period, and eccentricity of the outer companion in these CHT systems provides strong evidence of mutual orbit alignment, as well as a preference towards moderate outer eccentricities.}
   {Our novel approach has proven effective in identifying potential triple systems, thereby increasing their number in the catalogues. Our findings emphasise the importance of combined astrometric and RV data analysis in the study of multiple star systems.}


   \maketitle
%

\section{Introduction}
Triple star systems, where a distant tertiary star orbits an inner binary, offer fertile ground for advancing our understanding of stellar formation and evolution. These hierarchical systems provide valuable insight into the complex dynamics and interactions that can shape the life cycles of stars. Although binary stars have been extensively studied and serve as benchmarks for testing star formation theories, the exploration of triple systems has lagged behind due to the challenges inherent to their detection \citep{TokovininMSC97, Tokovinin04, Tokovinin14, Tokovinin17Alignment, Borkovits22}. This is particularly true for compact hierarchical triples (CHTs) for which the outer orbital period is less than $1000$ days, making them ideal cases for studying dynamical interactions within human-observable timescales.

The \gaia\  mission \citep{gaiamission16}, with its unprecedented astrometric precision, has revolutionised our ability to detect and characterise binary and multiple star systems. Among its many products, the third \gaia\  data release  \citep[DR3;][]{Vallenari23} contains the  Non-Single Stars (NSS) catalogue \citep[][]{NSS23}, which includes the orbital astrometric solutions for more than $169,000$ objects \citep{Halbwachs23}. 
However, the detection of CHTs, particularly those with unresolved inner binaries, remains limited. Most of the known compact hierarchical systems involve eclipsing binaries (EBs) detected by \textit{Kepler} \citep{Kepler10} and the Transiting Exoplanet Survey Satellite \citep[TESS;][]{TESS15}, whose outer companions are revealed through eclipse time variation or other effects \citep{Rappaport13, Borkovits16, Borkovits22, Rappaport22, Powell23, Mitnyan24}. 

Recently, \cite{Czavalinga23} cross-referenced over 1 million EBs with \textit{Gaia}'s NSS catalogue and identified $403$ potential triple systems, $376$ of which were newly discovered candidates. The study further validated $192$ candidates using TESS eclipse timing variations. Their results suggest that while \textit{Gaia}'s orbital periods are reliable, other orbital parameters should be interpreted with caution.

In this study we introduce a novel approach for finding potential CHT systems among \gaia\  astrometric orbital and accelerated solutions \citep{Halbwachs23}, by analysing the peak-to-peak radial velocity (RV) amplitude of available sources \citep{Katz23} and comparing it with the astrometric semi-amplitude. This method effectively identifies systems where the excess RV signal may indicate the gravitational influence of a third (inner) star, thereby separating likely triple systems from simple astrometric binaries.
By focusing on this unexplored aspect, our work aims to expand the content of the Multiple Star Catalogue \citep[MSC\footnote{http://www.ctio.noirlab.edu/~atokovin/stars/index.html};][]{TokovininMSC97, TokovininMSC18} by adding new compact hierarchies and to provide a more complete picture of their statistics. The limiting factor in our method is the fact that we are unable to find the orbital periods of the inner binaries and can only flag these sources as likely triple systems. 

The \gaia\  DR3 NSS sample of single-lined spectroscopic binaries \citep[SB1s;][]{Gosset24}, while helpful for cross-matching potential new candidates, is not a viable option for identifying hierarchical systems, as the \gaia\  pipeline assumes purely binary solutions. This limitation has resulted in numerous spurious orbits. A similar challenge arose in the refinement work of \cite{Bashi22}, where publicly available RVs from the Large Sky Area Multi-Object Fiber Spectroscopic Telescope \citep[LAMOST;][]{Cui12} and the GALactic Archaeology with HERMES  \citep[GALAH;][]{Buder21} were used to improve genuine SB1 detections by \gaia. Their analysis, while valuable, also assumed a binary star model, leading to misclassifications of triple systems. This issue is further highlighted in the work of \cite{Czavalinga23}, where many of the low-scoring sources, according to the criteria of \citet{Bashi22}, were in fact triple candidates, showing the limitations of current models to distinguish between binaries and triples.

\begin{figure*}
   \centering
\includegraphics[width=18cm] {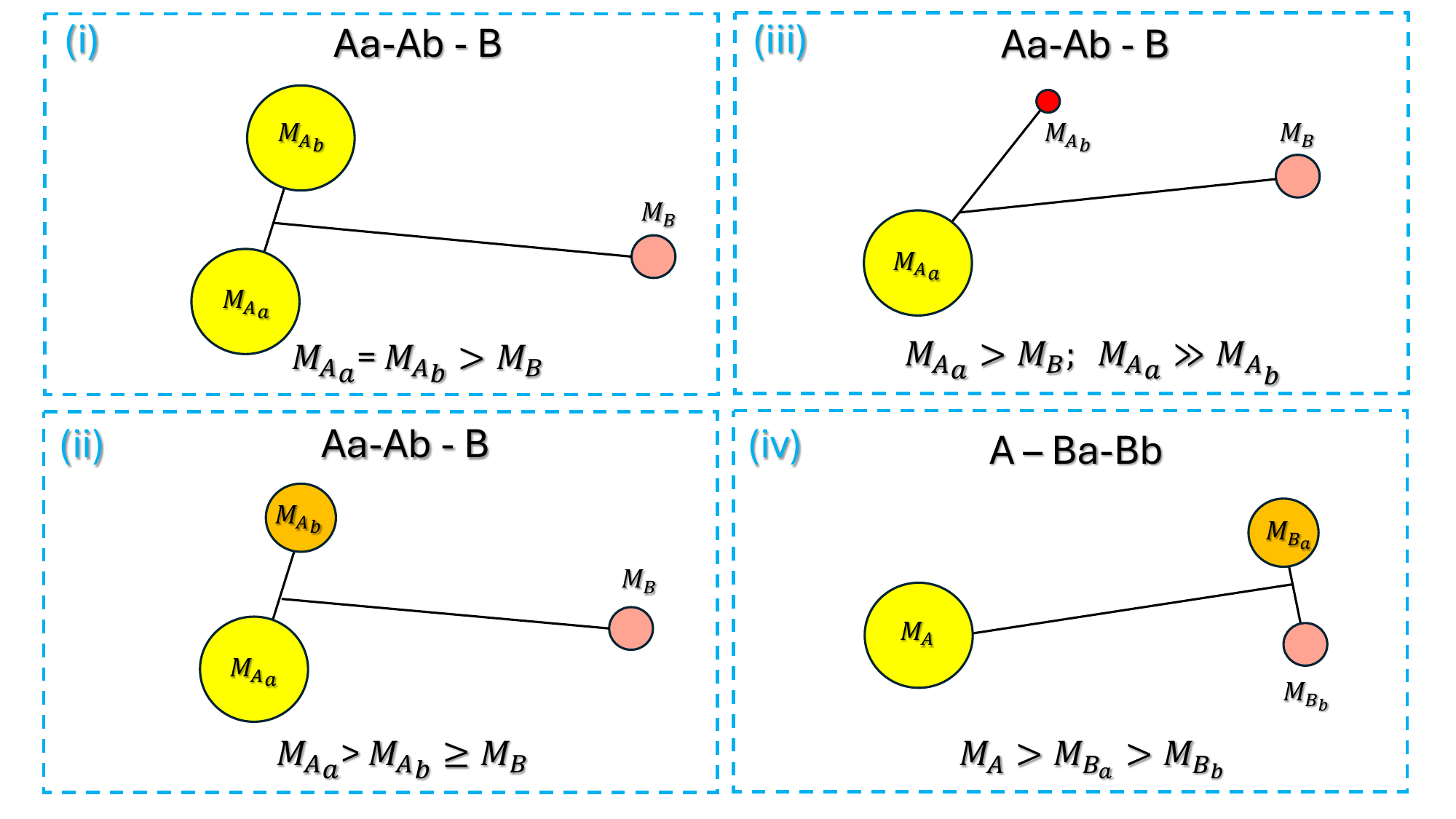}
   \caption{Schematic representation of the four possible configurations of triple star systems considered in this study. The first configuration (top left; Aa-Ab-B) consists of a close twin binary ($q_{\rm in}=M_{A_a}/M_{A_b} \simeq 1$) with an outer astrometric companion. In this case, the inner binary has nearly equal-mass stars, but the lower mass of the outer companion makes it more challenging to detect astrometrically. The second configuration (bottom left), to which our method is most sensitive, involves a close binary, $q_{\rm in} < 1,$ with an outer companion of lower mass. Here the inner binary’s spectroscopic signal is strong, while the outer companion dominates the astrometric signal. The third configuration (top right) is similar to the first two but with a significant mass ratio difference in the inner binary ($q_{\rm in} \ll 1$), where one star is much more massive than the other, making it difficult to detect its RV variation. Lastly, the fourth configuration (bottom right) represents a system where the primary (brightest) star is single, and the outer companion is a close binary system of lower mass (A-Ba-Bb) that contributes significantly to the astrometric signal. Each configuration presents unique challenges for detection and analysis based on the mass distribution and orbital architecture.}
              \label{fig:TripConfig}%
    \end{figure*}

Our method is well suited for detecting triple systems of Aa-Ab-B architecture, where the brightest component belongs to a close inner pair of stars with a moderate inner mass ratio, $q_{\rm in}$, accompanied by an outer astrometric companion. In this case, shown in the two left panels of Fig.~\ref{fig:TripConfig}, the inner binary causes a strong RV variation of the primary (brightest) star, while the outer companion produces the astrometric signal, making these systems ideal for our analysis. Our method is most effective when the inner binary has a mass ratio that is neither too small nor too large, such that the spectrum is dominated by the light of the brightest star and the \gaia\  RVs are those of this star. In cases where $q_{\rm in} \simeq 1$ (top left of Fig.~\ref{fig:TripConfig}), the inner binary consists of nearly equal-mass stars with opposite RV variations, making it more difficult to detect the system spectroscopically. Conversely, in systems where $q_{\rm in} \ll 1$ (top right of Fig.~\ref{fig:TripConfig}), our method encounters challenges due to the limited accuracy of \textit{Gaia}’s Radial Velocity Spectrometer (RVS) for detecting such low-mass companions. Similarly, in the A-Ba-Bb configuration (bottom right of Fig.~\ref{fig:TripConfig})), the RV variation of the brightest star reflects only motion in the astrometric orbit because stars in the close inner subsystem are much fainter and do not appear in the combined spectrum. 
Despite being difficult to detect through RVs, these systems can still have large astrometric amplitudes because the inner binary is more massive than a single star. Such configurations with astrometric mass ratios $q_{\rm out} > 1$ are expected to be more common than astrometric binaries with compact companions \citep[e.g. white dwarfs, neutron stars, or black holes;][]{Shahaf19, Shahaf23AMRF, Andrew22, El-Badry23, El-badry24}, which makes them important for understanding the broader population of triple stars. These diverse configurations illustrate how mass ratios and orbital architectures affect our method’s sensitivity to triple star systems.

Section~\ref{sec:Sample} details our sample selection process. We then split the analysis into two parts: CHT candidates, discussed in Sect.~\ref{sec:CHT}, and short-period binaries in accelerated solutions, covered in Sect.~\ref{sec:accel}. In each section we describe the methodology used, the validation process with the MSC, and how we utilised available RV epochs from publicly accessible ground-based spectrographs, as well as from EB and ellipsoidal variable TESS catalogues. Additionally, we highlight the key results from our exploration of potential CHT systems. Finally, Sect.~\ref{sec:discussion} provides a discussion of our findings and their implications for future research.

\section{Sample selection}
\label{sec:Sample}
We selected from the \gaia\  DR3 NSS catalogue \texttt{nss\_two\_body\_orbit} \citep{NSS23}, the fraction of orbital astrometric binaries using the condition \texttt{nss\_solution\_type = 'Orbital'}, yielding an initial sample of $169,227$ sources \citep{NSS23, Halbwachs23}. Similarly, to get the sample of accelerated sources, we used the \gaia\  acceleration catalogue, \texttt{nss\_acceleration\_astro} \citep{Halbwachs23}, which lists a total of $338,215$ sources.

We cross-matched these catalogues with the \gaia\  source catalogue \texttt{gaia\_source} to get RV information. We refined the sample by selecting sources with \texttt{rv\_method\_used = 1} as the RV amplitude information encapsulated in the \texttt{rv\_amplitude\_robust} field is only available for bright stars with $G_{\mathrm{RVS}} < 12$ \citep{Katz23,Sartoretti23}. 

The \texttt{rv\_amplitude\_robust} field provides the total amplitude in the RV time series, defined as the difference between the maximum and minimum robust RV values (maxRobust minus minRobust) after outlier removal. To identify these outliers, the \gaia\  team excluded valid transits with RV values outside the range Q1 - 3 × IQR and Q3 + 3 × IQR, where Q1 and Q3 are the 25th and 75th percentiles, and IQR = Q3 - Q1.

The \gaia\  RVS operates within a narrow spectral range of 846-870 nm. This specific band was chosen to optimise RV measurements for cooler stars, thereby limiting the ability to determine the velocities of hot stars. While \gaia\  DR3 includes a dedicated pipeline \citep{BlommeHotRV23} to assess the RVs of hot stars ($T_{\mathrm{eff}}=6900-14000$ K), the lower number of hot stars sources compared to cool stars, and the difference in pipeline assessment presents considerable challenges.
Consequently, we focused our analysis on the sample of cool stars using the criterion \texttt{rv\_template\_teff} > 3900 K and \texttt{rv\_template\_teff} < 6900. 

In addition, as the \gaia\  RVS is a slitless spectrograph, spectra of very
close sources might overlap and blend in crowded areas, especially towards the Galactic midplane, and bias the RV derivation. 
To avoid these rare cases, we filtered out sources where the fraction of de-blended spectra was larger than 0.5 (i.e. \texttt{rv\_nb\_deblended\_transits} / \texttt{rv\_nb\_transits} > 0.5). 

Lastly, to obtain a good phase coverage of the RV measurements, we followed \citep{Bashi24} and constrained our sample to sources with \texttt{rv\_visibility\_periods\_used} > 8. Consequently, we are left with a final orbital sample of $21,207$ sources and a final accelerated sample of $81,720$ sources.

\section{Triple candidates in astrometric solutions}
\label{sec:CHT}

We used the Thiele-Innes parameters of our orbital astrometric binaries and converted them into Campbell elements following \cite{Halbwachs23}. Using the photocentre angular semimajor axis, $a_0$, and astrometric parallax, $\varpi$, we estimated the linear astrometric semi-major axis $a_{\mathrm{phot}}=a_0/\varpi$ and the inclination $i$ of the astrometric orbits. Then
 we calculated the expected astrometric RV semi-amplitude $K_{\mathrm{0}}$ as
\begin{equation}
 K_{\mathrm{0}} = \frac{2\pi}{P} a_{\mathrm{phot}} \frac{\sin i}{\sqrt{1-e^2}},
  \label{eq:K_0}
\end{equation}
where $P$ and $e$ are the period and eccentricity of the astrometric orbit. 

To accurately determine the uncertainties $\sigma_{K_{\mathrm{0}}}$ associated with $K_{\mathrm{0}}$, we used a Monte Carlo approach. We utilised the correlation matrix listed in the NSS catalogue (\texttt{corr\_vec}) and randomly drew a set of $1000$ Campbell elements using a normal
distribution with the mean and standard deviation corresponding to the mean and uncertainty values reported for that system. In doing so, we accounted for the dependences between the parameters, allowing for a more robust estimation of the uncertainties in the  astrometric orbital elements. This approach ensures that the final uncertainties reflect both the intrinsic measurement errors and the propagation of those errors. As for the sampling of the eccentricity, similar to \cite{Bashi22}, we excluded cases with
negative or larger-than-one eccentricity values and used a circular mean and uncertainty when dealing with estimation of system's inclination.

To further exclude spurious astrometric solutions, we discarded sources with insignificant photocentre semi-amplitude using $a_{\mathrm{phot}}/\sigma_{a_{\mathrm{phot}}} < 5 $. This cut left us with $19,858$ orbital sources. 

As discussed in \cite{Andrew22}, \cite{Katz23}, and \cite{Bashi24}, the RV errors depend on the $G_{\mathrm{RVS}}$ magnitude, where fainter stars tend to show larger RV errors. 
This dependence is a critical issue that needs to be mitigated to ensure accurate estimations of $K^*_1$. 
For example, it can overestimate the RV-based semi-amplitude for fainter stars, suggesting systems containing these stars are more likely to exhibit higher apparent RV variations.
Consequently, this can artificially inflate the number of detected triple system candidates by introducing many false positives.

   \begin{figure*}
   \centering
    \includegraphics[width=0.45\textwidth] {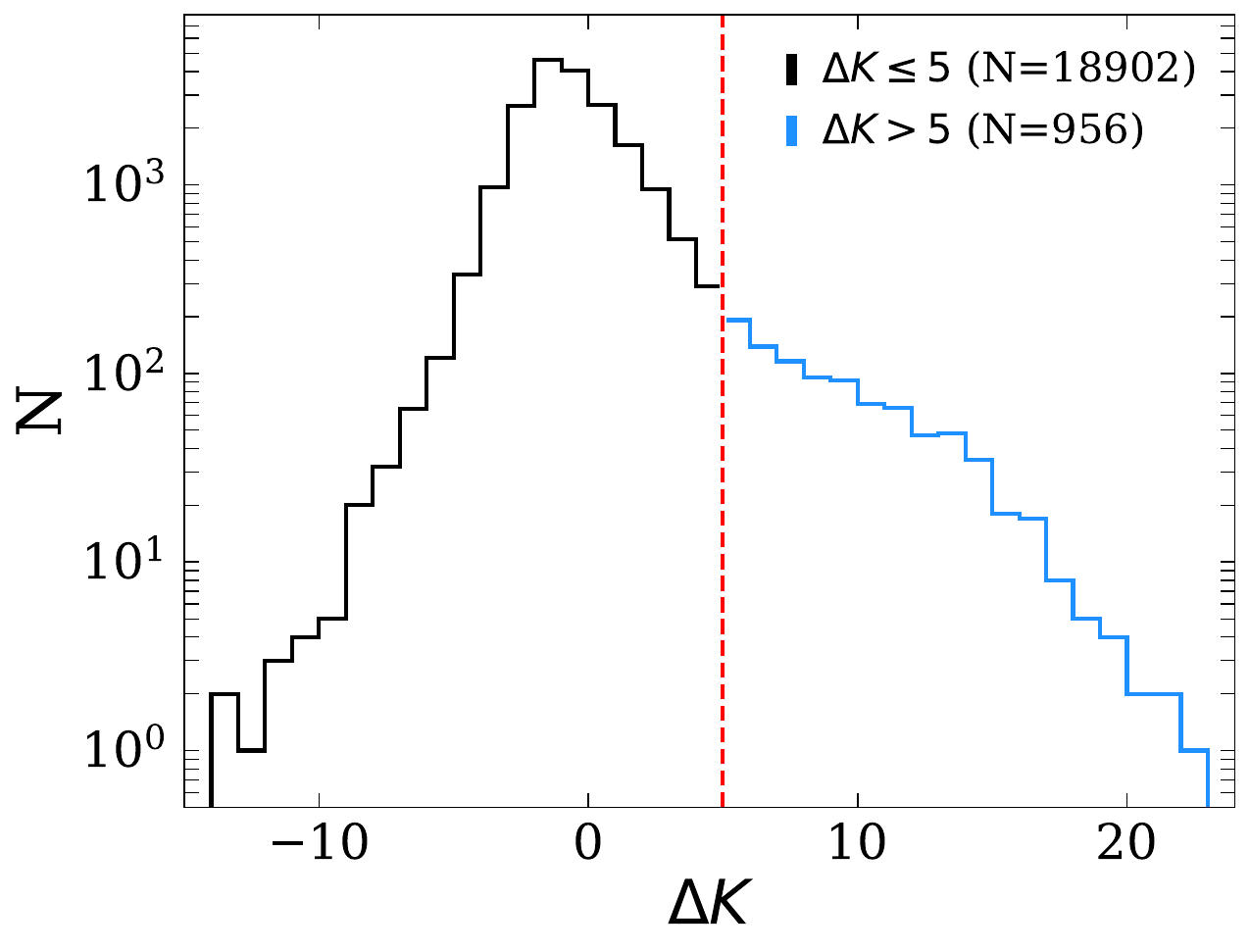}
    \includegraphics[width=0.45\textwidth] {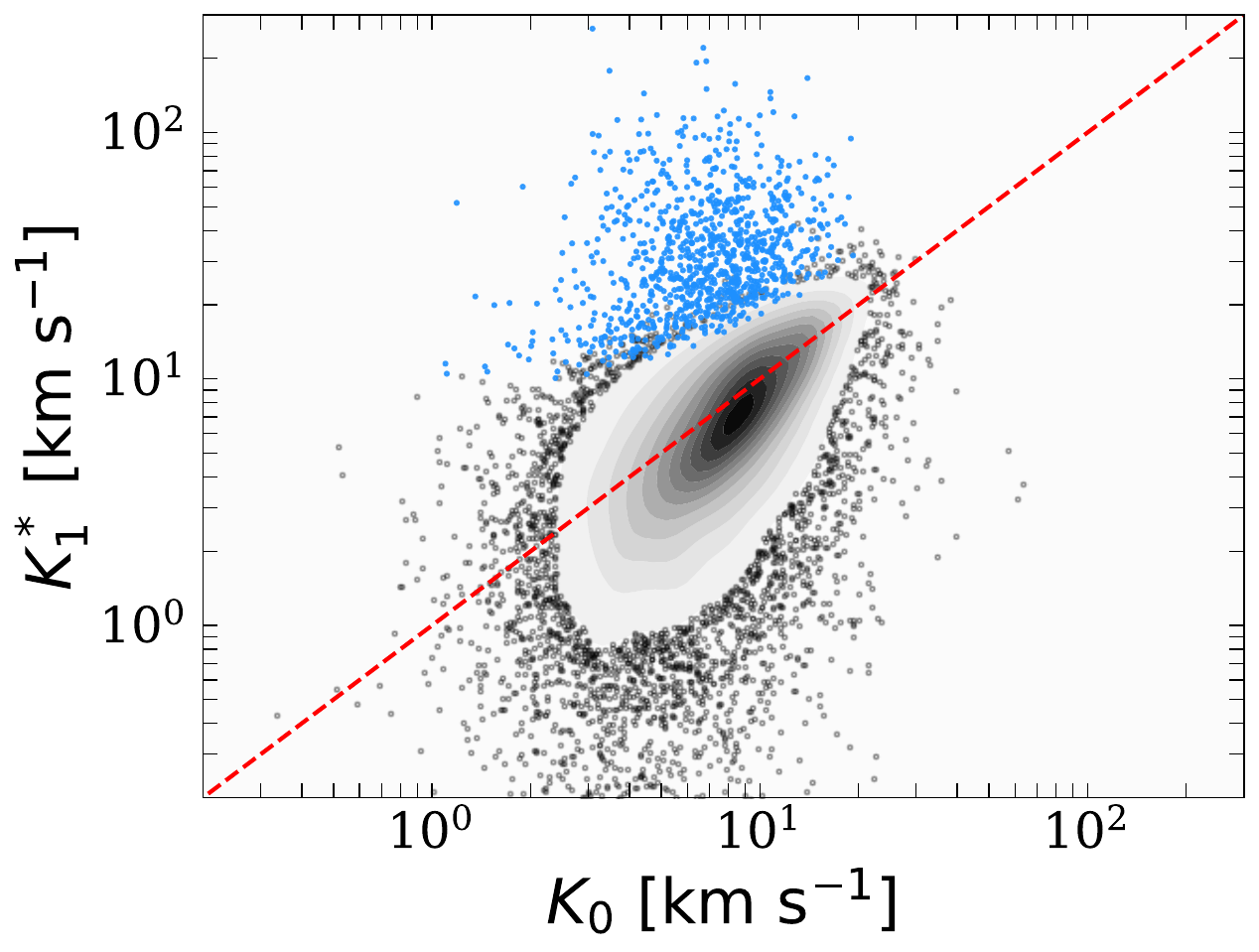}
   \caption{CHT candidates selection. Left panel: Histogram of $\Delta K$.\ The vertical dashed red line shows our threshold for selecting CHT candidate systems (marked in blue). Right panel: RV-based semi-amplitude as a function of the astrometric semi-amplitude $(K^*_1 - K_0)$ of the \textit{Gaia} astrometric binaries sample. Blue points mark sources with $\Delta K > 5$, which can be considered possible CHT systems (inner close binary and an outer astrometric binary), while black points and respective contours indicating regions of higher data concentration mark probable astrometric binaries with $\Delta K \leq 5$. The dashed red line marks a 1:1 relation.}
              \label{fig:K0_k1}%
    \end{figure*}

To mitigate this effect, we followed a similar approach as \cite{Bashi24}, using a Bayesian framework to model the population of sources in the $G_{\mathrm{RVS}}$ -- $\log \left(\mathrm{RV_{pp}}/2 \right)$ plane. A full description is provided in Appendix~\ref{Appendix: A}, but the main idea can be summarised as follows: the model characterises the single star population by fitting a Gaussian distribution with a mean trend $\mu\left ( G_{\mathrm{RVS}} \right )$, which captures the dependence of RV errors on the magnitude, and a standard deviation $\sigma_s$ that describes the scatter in RV errors for single stars of given magnitude. This approach helps us isolate \gaia\  systematics and distinguish between single stars and binaries. Given our fit (see the posterior values in Table~\ref{tab:priors} and the best fit model curve in Fig.~\ref{fig:RVamp_Grvs}), a typical single star with apparent RVS magnitude of $G_{\mathrm{RVS}} \simeq 12$ has a \gaia\  RV errors of $10 \pm 5$ \kms, while a $G_{\mathrm{RVS}} \simeq 6$ source has a $0.3 \pm 0.1$ \kms~RV variation.

Using this dependence, we approximated the RV-based semi-amplitude estimator of our full astrometric sample as
\begin{equation}
 K^*_1=\mathrm{RV}_{\mathrm{pp}}/2- 10^{ \mu \left(G_{\mathrm{RVS}}\right)} \,\cdot
 \label{eq:K*_1}
\end{equation}
The uncertainty $\sigma_{K^*_{\mathrm{1}}}$ primarily depends on $\sigma_s$, which is derived from the fit to the single-star model. Additionally, there is a dependence on the number of RV measurements, $N_{\mathrm{RV}}$, which, assuming a uniform distribution of the orbital phase, leads to the uncertainty of $\mathrm{RV}_{\mathrm{pp}}/2$ equal to its fraction $\sqrt{12 N_{\mathrm{RV}}}$. To account for potential underestimation of RV errors in bright sources, we included in quadrature an extra contribution of of $1$ \kms. The total uncertainty was then approximated as
\begin{equation}
 \sigma _{K^*_1}=\sqrt{\sigma_s ^2  + \left(\frac{\mathrm{RV}_{\mathrm{pp}}} {4 \sqrt{3 N_{\mathrm{RV}}}}\right)^2 + 1 }\,\cdot
\end{equation}

To convey the validity of this method in estimating the actual semi-amplitude of binary stars, we show in Appendix~\ref{Appendix: B} the results of a simulation to model how well the peak-to-peak RV variation represents the actual semi-amplitude $K_1$. Our simulations confirm the underestimation of the true peak-to-peak RV amplitude; thereby by using $\mathrm{ RV_{pp}}/2$ corrected for errors, we do not risk introducing false positives in our list of candidate triples. In addition, we show in Appendix~\ref{Appendix: B} the result of a cross-match between a sample of known binary stars listed in the Ninth Catalogue of Spectroscopic Binary Orbits \citep[$S_{\mathrm{B^9}}$;][]{sb9} with the \gaia\  catalogue. Overall, we find good agreement between the measured $S{_\mathrm{B^9}}$ $K_1$ and the \gaia\  semi-amplitude estimator. 

We then moved to estimate the difference between the estimated RV amplitude and the astrometric RV semi-amplitude in units of standard deviation using the expression
\begin{equation}
 \Delta K = \frac{K^*_1 - K_0}{\sqrt{\sigma_{K^*_{\mathrm{1}}}^2 + \sigma_{K_{\mathrm{0}}}^2}} \,\cdot
\end{equation}
Somewhat arbitrarily, we defined sources with $\Delta K > 5$ as possible hierarchical triple star systems, which yielded a subsample of $956$ candidates. 
A histogram of $\Delta K$ is plotted in the left panel of Fig.~\ref{fig:K0_k1}, showing our selection of CHT candidates in blue. As expected for a subsample of triple systems, a long tail in $\Delta K$ distribution is evident. We find that the median $\Delta K$ is $-0.77$, suggesting an underestimate of the true peak-to-peak RV amplitude, as expected by our simulations. 

To further convey our selection, we
show in the right panel of Fig.~\ref{fig:K0_k1} a scatter plot of $K^*_1$ versus. $K_0$. In this plot, systems with
$\Delta K > 5$ are marked with blue points. The CHT candidates tend to cluster on the upper side of the plot, where the RV-based semi-amplitude exceeds the astrometric semi-amplitude by a significant margin, as expected for triple systems with additional close inner pairs. In contrast, the majority of astrometric binaries, given their corresponding uncertainties, exhibit a closer agreement between their astrometric and RV-based semi-amplitudes, where the dotted red line marks a 1:1 ratio. 

 \subsection{Candidate quality assessment}

We find the distribution of sources across the Galactic plane to be consistent between the two groups, suggesting a similar spatial pattern. The \gaia\  scanning law’s non-uniformity might contribute to the observed clustering of sources in regions scanned more frequently. This similarity in the spatial distribution between our astrometric sample and the triple candidates suggests a potentially homogeneous selection effect due to \textit{Gaia}'s scanning patterns.

   \begin{figure*}
   \centering       \includegraphics[width=0.45\textwidth] {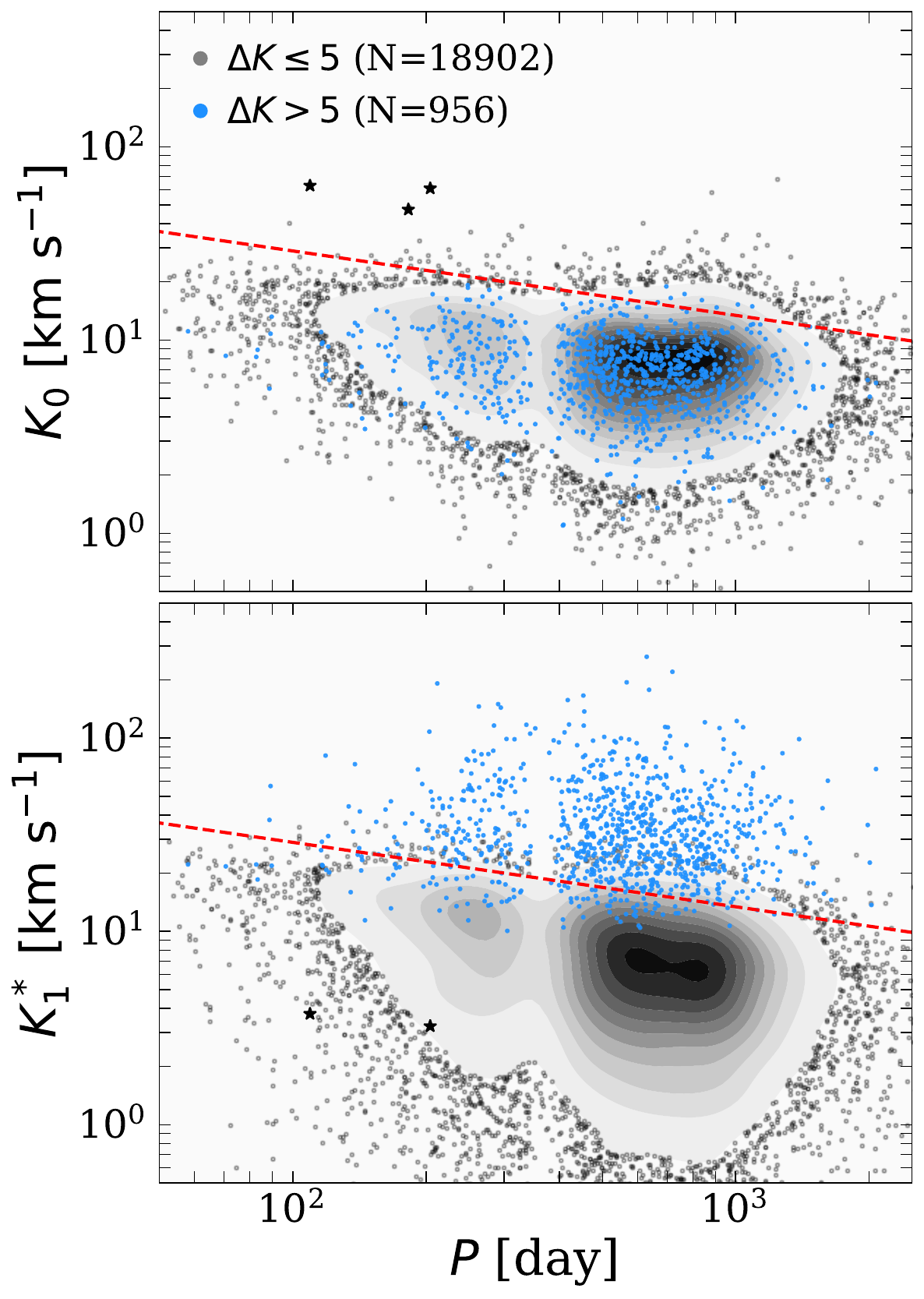}
\includegraphics[width=0.45\textwidth] {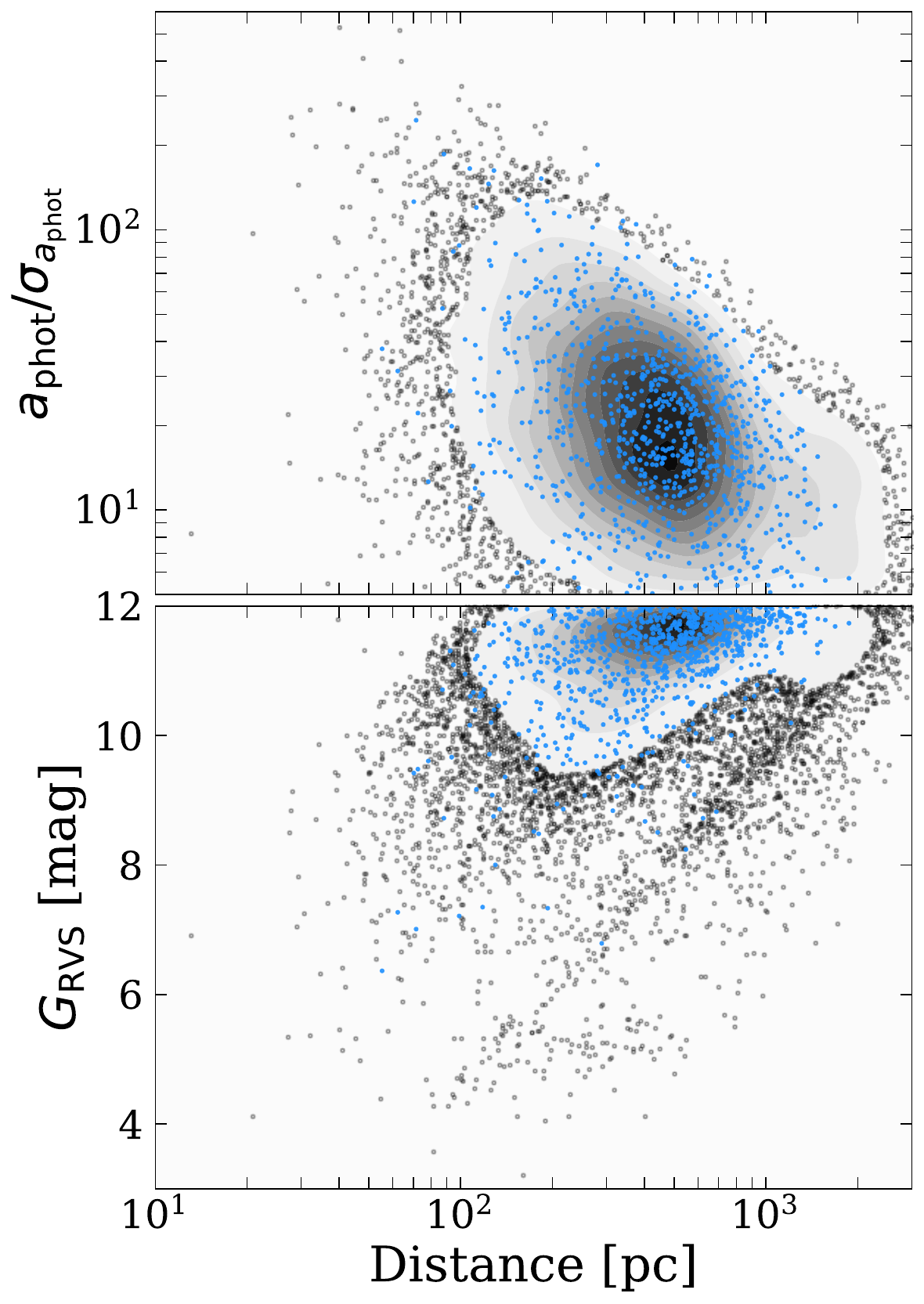}        
   \caption{Astrometric binaries (black) and CHT candidates (blue). Left panel: Astrometric RV semi-amplitude and the estimated RV amplitude as functions of the orbital period ,\( P \) $(P - K_0$ and $P - K^*_1)$. The dashed red lines indicate the expected semi-amplitude for an edge-on $1\;M_{\mathrm{\odot}}$ star with mass ratio $q = 1$ in a circular orbit. Black star symbols mark sources with $K_0$ significantly above the dashed line (see the main text). The vertical gap in the middle of the datasets reflects \textit{Gaia}’s insensitivity near 1-year-period binary orbits and its harmonics. Right panel: Photocentre significance (top panel) and apparent RVS magnitude (lower panel) as functions of distance $(\text{distance} - a_{\mathrm{phot}}/\sigma_{a_{\mathrm{phot}}}$ and $\text{distance} - G_{\mathrm{RVS}})$. }
              \label{fig:P_K0_K1}%
    \end{figure*}

On the left side of Fig.~\ref{fig:P_K0_K1} we show the distribution of $K_0$ and $K^*_1$ as a function of the astrometric orbital period, $P$. The $P-K_0$ diagram (top panel) shows a striking similarity between triples and binaries. Both populations seem to occupy similar regions of this parameter space. 
Given the expected semi-amplitude relation

\begin{equation}
 K = \left( \frac{2\pi G}{P}\right )^{1/3} \frac{M_1^{1/3} q \sin i}{(1+q)^{2/3}} \frac{1}{\sqrt{1-e^2}}, 
 \label{eq:K_1}
\end{equation}
where $M_1$ is the primary mass and $q=M_2/M_1$ is the outer mass ratio, we show by the dashed red lines in the left panels of Fig.~\ref{fig:P_K0_K1} the expected semi-amplitude as a function of orbital period for the case of an edge-on $1\;M_{\mathrm{\odot}}$ solar mass star with $q=1$ in circular orbit. Following Eq.~\ref{eq:K_1}, a $K\sim P^{-1/3}$ trend is expected, and indeed, our sample does follow this trend rather nicely. Black points above the dashed line can be triple systems with $q_{\rm out|} > 1$ where the close pair is hosted in the secondary component of astrometric pairs; such triples are not detectable by our method. 

The RV amplitude of an astrometric binary is estimated by Eq. \ref{eq:K_0} assuming that the astrometric amplitude $a_1$ describes motion of the primary component. When the light of the secondary component is non-negligible, the amplitude of the photocentre motion $a_{\mathrm phot}$ is reduced by the factor
%
\begin{equation}
 \frac{a_{\mathrm{phot}}}{a_1}  =  1 - \frac{S(1+q)}{q(1+S)} ,
 \label{eq:a_phot}
\end{equation}
where $S=I_2/I_1$ is the flux ratio in the $G$ band $q$ is the outer mass ratio. For nearly equal components with $S \approx 1$, $a_{\mathrm phot}$ tends to zero; 
the amplitude reduction by blending is notable for astrometric binaries with $q >0.5$. Such binaries are situated well below the dashed red line. Blending also reduces their RV amplitude $K_1^*$. 
 
On the other hand, systems with compact astrometric companions such as white dwarfs, neutron stars, or black holes, where $q > 1$ and $S \approx 0$, can appear above this line since the compact object does not contribute light, causing the astrometric and RV-based semi-amplitude to be identical \citep{Shahaf19, Andrew22, El-Badry23, Shahaf23AMRF, El-badry24}. However, we find most sources above the dashed red line are consistent with the general trend when uncertainties in $K_0$ are considered. Three notable cases, marked by black star symbols that are significantly ($>3\sigma$) above the dashed red lines, illustrate additional complexities. Gaia DR3 4482912934572480384, originally listed as a neutron star candidate \citep{Shahaf23AMRF} with $P = 182$ days, exhibits a much lower $K^*_1$ value than predicted ($K_0 = 47$\kms, $K^*_1 = 0.067$\kms). However, \cite{El-badry24} excluded this source from the compact object status after performing extended RV follow-ups, suggesting that applying our RV method earlier could have saved observational resources. Gaia DR3 3509370326763016704 ($P = 109$ days, $K_0 = 63$\kms, $K^*_1 = 3.7$\kms) followed a similar pattern \citep{Shahaf23AMRF} and was excluded from the neutron star candidate sample after RV measurements \citep{El-Badry23}. Finally, Gaia DR3 5152756278867291392 ($P = 204$ days), identified as an EB with a period of 2.7 days, presents an unusual case where $K_0 = 61$\kms~far exceeds $K^*_1 = 3$\kms. This discrepancy, combined with a high eccentricity ($e = 0.88$), needs further investigation.
These cases, though few in number, illustrate the need to account for the influence of \textit{Gaia}’s yearly scanning cycle on systems with periods between 109 and 204 days.


In contrast, in the $P-K^*_1$ diagram, the triples are predominantly located at the high end of the distribution above the dashed red line. This suggests that the RV-based semi-amplitude measured by \gaia\ is actually the semi-amplitude of the inner compact binary in a triple system, not the semi-amplitude of the outer astrometric binary.

The right side of Fig.~\ref{fig:P_K0_K1} compares the sample of astrometric binaries and triple candidates' photocentre significance $a_{\mathrm{phot}}/\sigma_{a_{\mathrm{phot}}}$ (top panel) and apparent RVS magnitude $G_{\mathrm{RVS}}$ (lower panel) as a function of the system distance. The median distance of the orbital sample is $410$ pc, while that of the CHT candidates is $460$ pc. The larger distance is expected given the larger mass and greater luminosity of these systems. A higher significance is expected in systems that are closer as well as brighter. In general, the CHT candidates do not populate a phase space different from the overall sample. The distinct sequence of brighter stars in the bottom panel is composed mainly of giant stars and does not include many triple candidates.

A partial list of all astrometric sources, including our $\Delta K$ metric used to select the CHT candidates, is given in Table~\ref{tab:DeltaK_list}. The full table is available online.

\begin{table}
        \centering
        \caption{ \gaia \ orbital source IDs sorted by $\Delta K$ (extract).}
 \small
        \label{tab:DeltaK_list}
        \begin{tabular}{cccc} 
                \hline \hline
                Gaia source id  & $K_0$ & $K^*_1$ & $\Delta K$ \\

                  & [\kms] & [\kms] &  \\
                \hline
1275790936877524608 & $11.64 \pm 0.62$ & $76.60 \pm 2.88$ & $22.07$ \\
1281813580534856448 & $3.47 \pm 0.11$ & $177.97 \pm 8.11$ & $21.51$ \\
5060565939731334400 & $4.42 \pm 0.14$ & $83.97 \pm 3.73$ & $21.30$ \\
5009772385177638144 & $4.85 \pm 1.03$ & $117.71 \pm 5.43$ & $20.44$ \\
2926655621040197248 & $6.00 \pm 0.65$ & $77.12 \pm 3.45$ & $20.25$ \\
2964349937657635072 & $3.76 \pm 0.08$ & $54.57 \pm 2.61$ & $19.45$ \\
1458572925243347584 & $10.76 \pm 0.77$ & $145.88 \pm 6.99$ & $19.22$ \\
5393359105543586304 & $5.92 \pm 0.35$ & $105.52 \pm 5.19$ & $19.15$ \\
4873737400679728256 & $3.08 \pm 0.09$ & $263.80 \pm 13.66$ & $19.08$ \\
842444370389914752 & $6.40 \pm 0.88$ & $191.96 \pm 9.73$ & $18.99$ \\
                \hline

\multicolumn{4}{l}{\footnotesize Note: The full table is available at the CDS.} \\
        \end{tabular}
\end{table}

\subsection{Validation}
\label{sec:CHT_validation}
In our efforts to validate the new CHT candidates, we conducted a cross-match with other relevant astronomical catalogues. Mainly, we looked for common sources reported in the MSC {\citep{TokovininMSC97, TokovininMSC18} or compared our sample with the recently published catalogue of EBs known to be part of a triple system \citep{Czavalinga23}. Additionally, we looked for common sources listed in the catalogue of TESS ellipsoidal variables \citep{Green23}. 

The comparison with these catalogue is informative as it allows us to check for consistency between the CHT candidates and well-documented triple systems with an inner compact binary. Similarly, the ellipsoidal variable catalogue provides an additional layer of validation, as these stars exhibit characteristic brightness variations due to their distorted shapes in close binary systems, regardless of the geometrical line-of-sight configuration required by EBs systems.

In addition, since \gaia\  DR3 does not include individual RV epochs, we used available ground-based spectroscopic surveys, specifically LAMOST \citep{Cui12}, GALAH \citep{GALAH15}, and the Apache Point Observatory Galactic Evolution Experiment \citep[APOGEE;][]{MajewskiAPOGEE17}, to explore detailed RV variations over time. For two promising cases with high $\Delta K$, we also performed follow-up RV monitoring with CHIRON \citep{CHIRON13}.  

\subsubsection{Cross-match with the Multiple Star Catalogue}

We ran a sky cross-match with a maximum angular distance of 1 arcsec between the $905$ compact triple candidate sample and the $12,484$ stars listed in the \texttt{component} table of the MSC \citep{TokovininMSC97, TokovininMSC18}, using its latest online version (December 2023).

This cross-match yielded a subsample of $37$ common sources, with most of them (34) overlapping with the  \cite{Czavalinga23} catalogue (i.e. they have an inner EB and an outer astrometric binary, which is discussed in the following).
In some systems, MSC lists additional resolved companions, meaning that they are quadruples of 3+1 hierarchy. The system 12222$-$2413 is a compact 2+2 quadruple containing two EBs \citep{Kostov22} The system 00514$-$3832 is not known to be eclipsing, but its large RV variation was established by \cite{Nordstr04} The $637$-day astrometric orbit in 14501+2939 is in tension with the eclipse time variation, which indicates a much longer period of 3493 days \citep{Barani15}

\subsubsection{Inner eclipsing binaries in the triple system catalogue}
We cross-matched our candidate list with the results presented in the recent study of \cite{Czavalinga23} focused on identifying CHT systems among \gaia\  astrometric binaries. They used several EB catalogues from various publicly available sky surveys, encompassing over a million targets, to search for \gaia\  DR3 NSS orbital solutions indicative of tertiary stars with orbital periods significantly longer than the eclipse periods. \cite{Czavalinga23} found $403$ objects with suitable \gaia\  orbital solutions, including $376$ new hierarchical triple system candidates. 

By performing a cross-match with this dataset, we found $59$ sources that overlapped with our sample, of which $43$ were part of our triple subsample, suggesting a $\sim 73\%$ fraction of common sources. We show in Fig.~\ref{fig:hist_dK_EB} a histogram (green) in $\Delta K$ of the cross-match between our triple candidates and the EB sample. The grey background histogram shows our overall sample of Orbital sources.
This significant overlap not only supports the validity of our triple-star candidates but also highlights the effectiveness of our methodology in detecting hierarchical triple systems. It is possible, at least in principle, that some associations between \gaia \ astrometric binaries and
eclipsing subsystems found by \cite{Czavalinga23} are spurious: the EB could be an unrelated background source causing flux variability in the large TESS pixels. These false positives are more common in crowded areas of the sky. 

To investigate this possibility further, we examined some of the reported EBs of \cite{Czavalinga23} with very low $\Delta K$ values. Two examples are TIC 158579379 (Gaia DR3 1617361920624734464) with $\Delta K = -5.69$, and TIC 76989773 (Gaia DR3 6551292978718711168) with $\Delta K = -0.83$. In the case of TIC 158579379, both primary and secondary eclipses are evident in the light curve; however, given the normalised flux depths (primary: $\sim 1.5 \%$; secondary: $\sim 0.5\%$) and shapes, this is likely a background EB as such shallow eclipses with these depths and shapes are characteristic of blended light from an unrelated EB. Indeed, using the \gaia \ archive, we found a nearby (within 15 arcsec) faint source (Gaia DR3 1617361916328722432, $G=18.38$ mag) that might be an EB system. 
For TIC 76989773, only the primary eclipse is apparent (normalised depth $\sim 0.5\%$), which could originate from a nearby (within 10 arcsec) comparable magnitude source (\gaia DR3 6551292974423994624, $G=11.6$ mag). In both cases, eclipses from nearby background sources could be misattributed to the target star because the large TESS pixel size of approximately $20$ arcsec \cite{TESS15} allows light from nearby EBs to contaminate the observed light curve.

On the other hand, because our method of selecting triple systems is less sensitive to systems of the A-Ba-Bb architecture compared to the eclipsing method, the presence of candidates with $\Delta K < 5$ in their sample is expected. Nevertheless, the overall consistency between our findings and those of the \cite{Czavalinga23} study strengthens the credibility of our sample, particularly in detecting new triple systems that do not eclipse.

   \begin{figure}
   \centering
\includegraphics[width=0.5\textwidth] {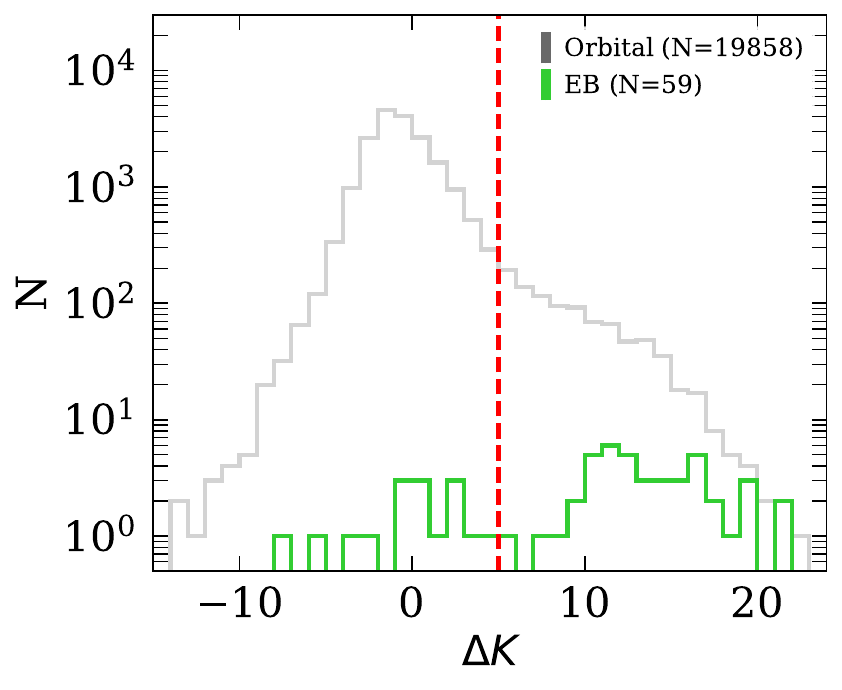}
   \caption{Histogram of $\Delta K$ of the full orbital sample (grey) and a subsample of triple systems with  inner EB pairs in common with \cite{Czavalinga23} (green). The vertical dashed red line shows our threshold for selecting CHT candidate systems.}
              \label{fig:hist_dK_EB}%
    \end{figure}

\subsubsection{Ellipsoidal catalogue}
We also performed a cross-match with \cite{Green23} focused on candidate binary systems exhibiting tidally induced ellipsoidal modulation, selected from TESS full-frame image light curves. \cite{Green23} have identified $15,779$ candidate binaries with main sequence primary stars and orbital periods shorter than $5$ days. 

Our cross-match with this ellipsoidal binary catalogue resulted in nine overlapping sources, of which six ($67\%$) were part of the CHT subsample. While the overall cross-match was rather low, as most of the sources in the ellipsoidal catalogue were part of hotter ($T_{\mathrm{eff}} > 6500$ K) systems, the overlap of sources with $\Delta K >5$ further reinforces the validity of our triple candidates, particularly in relation to systems with short-period binaries where the presence of a third body may significantly influence the observed dynamics. The consistency between our findings and those reported in this study provides additional support for the accuracy of our triple-star identifications, especially in systems where tidal forces and tertiary companions play a significant role.
The relatively low fraction of common triple sources, compared to the EB fraction, can be attributed to the fact that ellipsoidal binaries are not limited to line-of-sight geometries. As a result, we expect to detect them in cases with lower semi-amplitude values, where the RV-based semi-amplitude is smaller. This geometrical factor likely contributes to the reduced overlap, as systems with lower RV amplitudes might still exhibit significant ellipsoidal variations detectable photometrically.

   \begin{figure}
   \centering
         \includegraphics[width=0.5\textwidth] {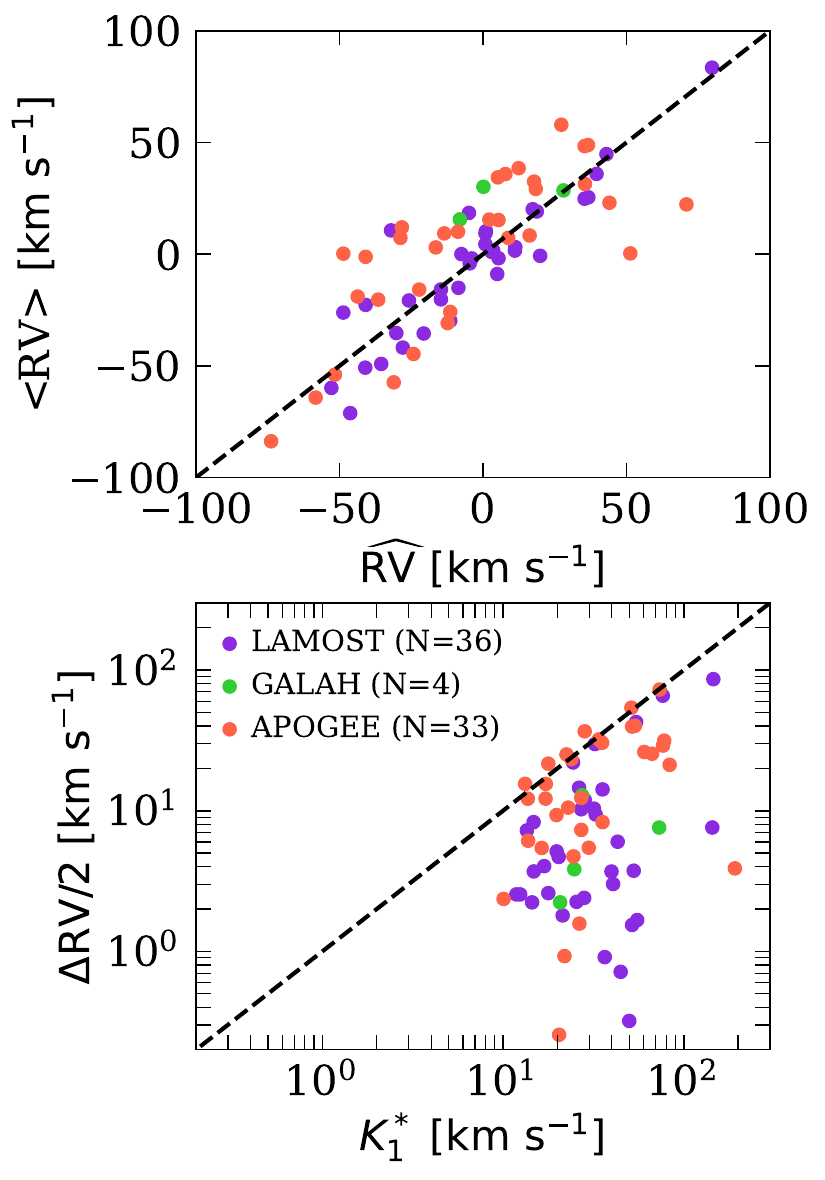}

   \caption{Ground-based RVs validation. Upper panel: Scatter plot of the average surveys RV epochs as a function of the median \textit{Gaia} RV value $(\widehat{\mathrm{RV}} - \left\langle\mathrm{RV}\right\rangle)$. Lower panel: Half the peak-to-peak variation in the surveys' RV epochs as a function of the estimated spectroscopic semi-amplitude $(K^*_1 - \Delta \mathrm{RV} / 2)$. Violet points mark sources observed by LAMOST, green points sources observed by GALAH, and red points sources observed by APOGEE. Dashed black lines mark a 1:1 relation.}
    \label{fig:RV_validation}%
    \end{figure}

\subsubsection{Ground-based RVs}
By utilising the RV data from ground-based surveys, we aimed to track the orbital motion of triple stars candidates ($\Delta K >5$), which can help us further constrain the orbital parameters of the systems. This complementary RV information is particularly useful for confirming the presence of the inner binaries in the triple star candidates, as it allowed us to detect short-term velocity trends that were otherwise not released as part \gaia\  DR3. 

We used the LAMOST DR6 \citep{Cui12}, GALAH DR3 \citep{Buder21}, and APOGEE DR17 \citep{APOGEE17} catalogues and cross-matched them with our orbital sample leaving sources with multiple epochs of RV measurements ($N_{\mathrm{RV}} >1$). The cross-match of our triple candidates with LAMOST DR6 yield $36$ sources, $28$ of them have two RV measurements. 
The cross-match of the triple candidates with GALAH DR3 yielded 4 sources, all of them have two RV measurements.
Lastly, the cross-match with APOGEE DR17, yielded the largest sample of sources with more than two RV epochs with $26$ out of $33$.

We list in Table~\ref{tab:groundRV} the cross-match results with the various ground-based surveys. The table includes essential parameters used to assess the RV variability and to characterise the triple star systems. Each row corresponds to a unique \gaia DR3 source \texttt{source\_id}, median \gaia RV $\widehat{\mathrm{RV}}$, astrometric RV semi-amplitude of the outer companion $K_0$, and the estimated spectroscopic semi-amplitude $K^*_1$. From the ground-based instrument side, the table details the number of RV measurements, $N_{\mathrm{RV}}$, obtained, along with the time span $\Delta T$ in days over which the RV data were collected. Additionally, $\left\langle\mathrm{RV}\right\rangle$ denotes the mean RV for the source over the measurement period, and $\Delta \mathrm{RV}$ represents the peak-to-peak variation in the RVs, providing a measure of the overall RV variability observed for each source.

The upper panel of Fig.~\ref{fig:RV_validation} presents a comparison between the median \gaia\  RV, the $\widehat{\mathrm{RV}}$ and the mean RV epochs, $\left\langle\mathrm{RV}\right\rangle$ of the three different surveys: LAMOST (violet), GALAH (green), and APOGEE (red). The scatter plot shows a clear correlation. In the lower panel we explore the relationship between the \gaia-estimated spectroscopic semi-amplitude $K^*_1$, and half of the maximum RV difference $\Delta \mathrm{RV} / 2$ given the surveys' RV epochs. An upper envelope is seen between $K^*_1$ and the RV difference, indicating that systems with higher RV variability tend to exhibit larger semi-amplitudes. The consistently higher values of $K^*_1$ across all cases suggest that the limited number of epochs available from the surveys may contribute to the more pronounced discrepancies, particularly in cases where the number of RV measurements, $N_{\mathrm{RV}}$, is low. The separation of data points by colour allows us to distinguish the surveys and highlights that, despite differences in instrumentation and observing strategies, the fundamental dynamical properties of the systems are similarly captured by each survey. 

We checked two other triple candidates with large $\Delta K$ by taking spectra with CHIRON \citep{CHIRON13} during several nights. The spectral resolution was 28\,000. To determine the RVs, spectra were cross-correlated with a binary mask based on the solar spectrum. The results are given in Table~\ref{tab:chiron}. The first star, \gaia \ DR3 4989698670108892288 (TYC 7001-1442-1), has a very wide ($\sim$50 \kms) and shallow correlation dip, so the RVs are measured with large errors. Their peak-to-peak scatter of 104 \kms~agrees with the large $K_1^* = 85$ \kms~estimated by \gaia. The large RV amplitude and broad lines suggest that this is might be a contact pair with  a period of a few hours.  However, the light curves in the  TESS sectors 29  and 30 do  not show any  significant periodic modulation, and the rms fluctuations of the normalised flux are only 0.4\%, so this pair is not eclipsing.  

The RVs of the second candidate, \gaia\ 
DR3 5050923154035615872 (TYC 7022-556-1), are measured reliably (its dip has rms width of 15\kms) and also show obvious variability. The six RVs match a circular orbit with a period of 2.2079 days, $T_0 = 2460553.32$ (RV maximum), $K_1$ of 42.0\kms, and centre-of-mass velocity of 25.2\kms. The dip width corresponds to a synchronous rotation. The \gaia-based estimate of the RV amplitude $K_1^*=41$ \kms~agrees well with this orbit, but the mean \gaia\  RV of $-31.73$ \kms~is in strong tension (the RV amplitude in the outer orbit $K_0 = 12.5$\kms is too small to explain the difference). 


\begin{table}
        \centering
        \caption{RVs of two candidates measured with CHIRON.}
        \label{tab:chiron}
        \begin{tabular}{lc} 
 \multicolumn{2}{c}{\gaia DR3 4989698670108892288 } \\
 
                \hline \hline
                JD & RV  \\
               &  {[\kms]} \\
                \hline
 2460553.7713 &  75.671   \\
 2460558.7861 &  47.871  \\
 2460562.7642 & $-$28.168  \\
                \hline
        \end{tabular}

        \vspace{0.5cm} 

        \begin{tabular}{lc} 
 \multicolumn{2}{c}{\gaia DR3 5050923154035615872 } \\
                \hline \hline
                JD & RV  \\
               &  {[\kms]} \\
                \hline
  
2460553.8695 & 12.070  \\
2460556.8730 & 63.304  \\ 
2460561.7601 & 19.232  \\ 
2460580.7922 & 60.084  \\ 
2460582.7901 & 36.378  \\
2460586.7820 & $-$7.409 \\
                \hline
        \end{tabular}
\end{table}


%
%
\subsection{Astrophysical properties}
\label{sec:CHT_results}

In this section we explore several astrophysical features of the CHT candidate sample and compare their orbital properties with those of the rest of the astrometric binary sample.

   \begin{figure}
   \centering
   \includegraphics[width=0.45\textwidth] {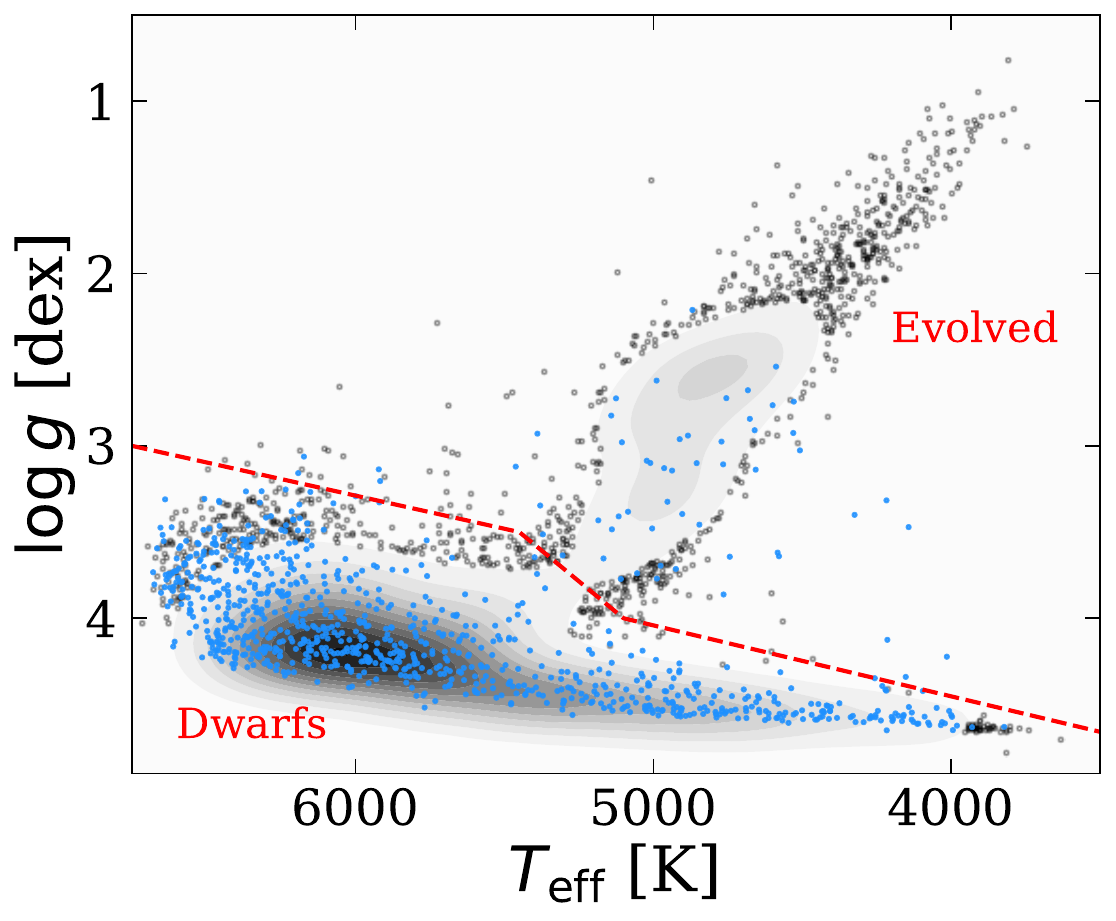}

   \caption{Effective temperature, \( T_{\mathrm{eff}}, \) as a function of the surface gravity $(\log g - T_{\mathrm{eff}})$ of astrometric binaries (black points and respective contours indicating regions of higher data concentration) and CHT candidates (blue points). The dashed red line delineates the division between dwarf and evolved systems.} 
              \label{fig:kiel}%
    \end{figure}

We started by exploring the distribution of astrometric binaries using a Kiel diagram of stellar gravities, $\log g$, as a function of effective temperature, $T_{\mathrm{eff}}$, based on the reported values derived in \cite{Andrae23}. We show in Fig.~\ref{fig:kiel} a scatter plot of the orbital sample, with blue points marking triple star candidates. The dashed red line divides our sample into dwarfs and evolved stars. Using this cut, $14,603$ binaries and $871$ triple candidates are defined as dwarfs and $4,299$ binaries and $85$ triple candidates are evolved. In particular, we find the ratio of the triple fraction in the dwarf to evolved subgroup to be of the order of $3$, which is in line with expectations, as CHT systems, and close binary systems in general, should not survive phases of stellar evolution where common envelope and mass transfer commence \citep{Toonen20}. 

In what follows, we use only dwarf systems in exploring the orbital features of the astrometric binary and CHT subsamples. By focusing on dwarf stars, we avoided potential contamination from stellar evolution effects, which could otherwise influence the observed orbital parameters and lead to biased interpretations of the systems’ dynamical properties. 

The mutual alignment between the outer and inner orbits in triple stars provides important clues about the formation and dynamical evolution of these systems. Understanding the relative orientation of these orbits -- whether they are co-planar or exhibit significant misalignment, can shed light on the processes that led to the current configurations. We analysed the distribution of the cosine of the inclination angle ($\cos i$) for our sample of astrometric binaries. The histogram of $\cos i$ in Fig.~\ref{fig:inc} shows a significant preference for face-on configurations (i.e. $\cos i \approx \pm 1$), which is expected given the bias in detecting astrometric binaries. Astrometric observations are more sensitive to systems with face-on orbits, as the apparent motion of the photocentre is maximised in these configurations. In a sample with isotropic orbit orientation, we expect a uniform $\cos i$ distribution. 

However, when we examine the $\cos i$ distribution of our triple candidates, we note a larger fraction of sources with $\cos i \approx 0$, indicating that their astrometric orbits have preference of edge-on orientation. This makes sense from a spectroscopic perspective, as the RV semi-amplitude ($K_1$) is maximised in edge-on orbits, making them easier to detect via RV measurements. 
The RV variation in our triple-star candidates is attributed to motion in the inner orbits. The distribution of $\cos i$ in the outer (astrometric) orbits provides evidence of some mutual orbit alignment. It is difficult to quantify the degree of such alignment without detailed modelling of all selection effects. 

   \begin{figure}
   \centering
   \includegraphics[width=0.5\textwidth] {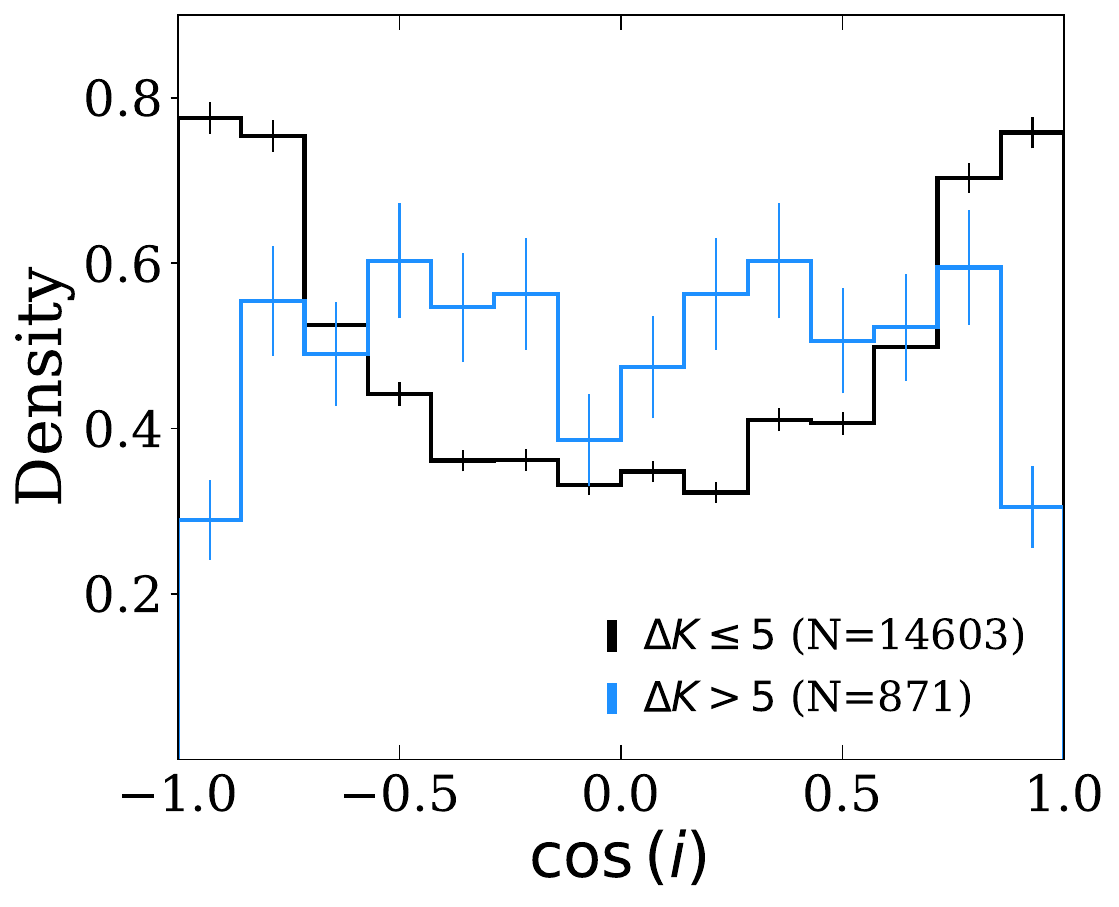}

   \caption{Histogram of $\cos i$ for astrometric binaries (black) and CHT candidates (blue). Vertical error bars represent Poisson uncertainties in each bin, calculated as $\sqrt{N_{\text{bin}}}$. Only dwarf stars based on the selection shown in Fig.~\ref{fig:kiel} are considered.} 
              \label{fig:inc}%
    \end{figure}

   \begin{figure*}
   \centering
\includegraphics[width=18cm]{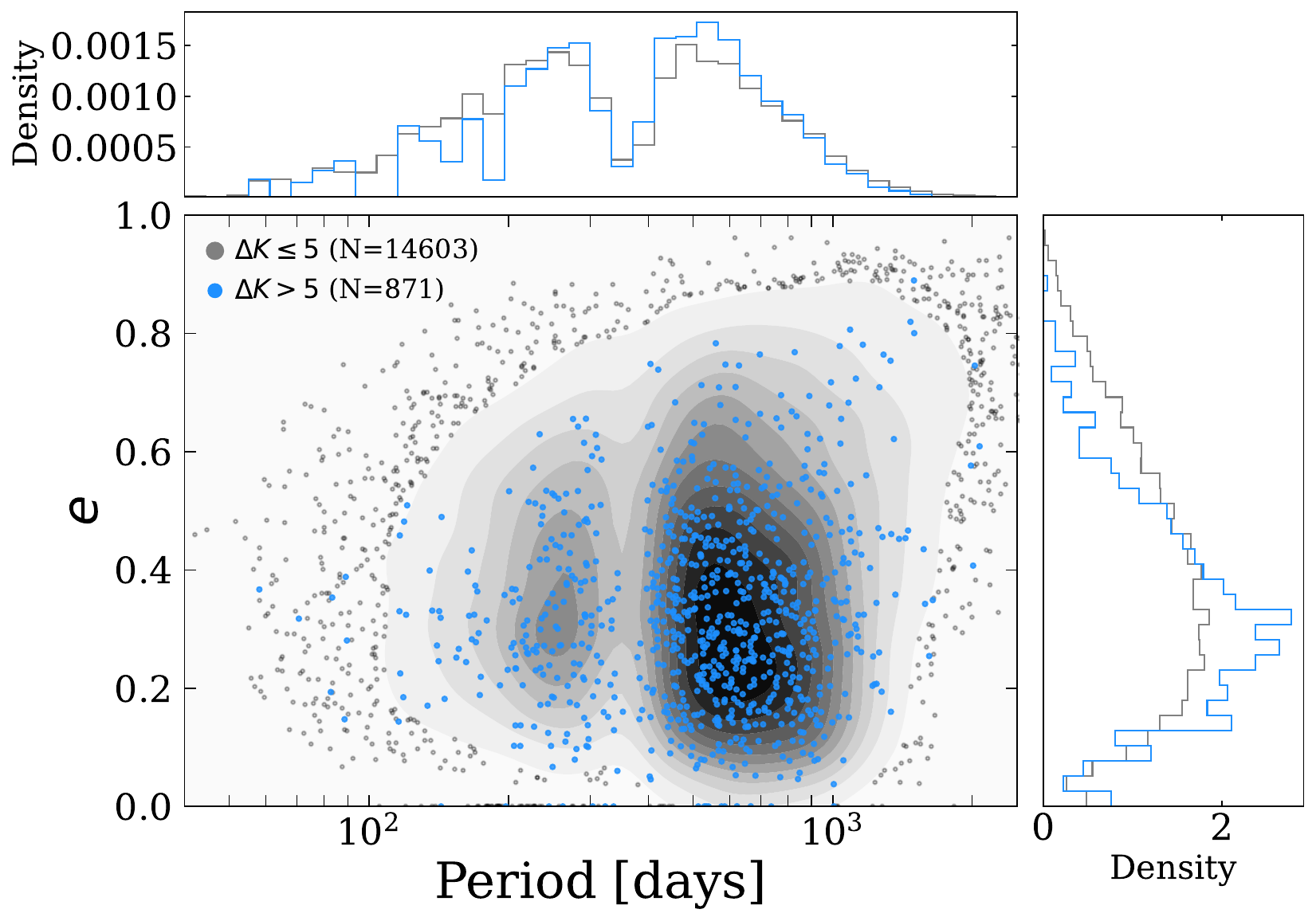} 
\caption{Period-eccentricity ($P-e$) diagram of astrometric binaries (black points and respective contours indicating regions of higher data concentration) and CHT candidates (blue points), with marginal distributions of period and eccentricity on the sides. Only dwarf stars based on the selection shown in Fig.~\ref{fig:kiel} are considered.} 
              \label{fig:per_ecc}%
    \end{figure*}

Next, we aimed to follow the distribution of sources on the period-eccentricity ($P-e$) diagram. The importance of the $P-e$ diagram for binary populations lies in the insights it provides into the formation, evolution, and dynamical interactions of these systems \citep{Bashi23}. We show in Fig.~\ref{fig:per_ecc} the period-eccentricity scatter plot of the orbital binaries (black points) and CHT candidates (blue points), with marginal distributions of period and eccentricity on the sides. 

The orbital period of the outer star in compact triples is a crucial factor in ensuring dynamical stability. Typically, hierarchical triple systems remain stable when the outer star has a significantly longer period $P_{\rm out}$ than the inner binary $P_{\rm in}$ to avoid strong gravitational interactions that could destabilise the system \citep{EggletonKiseleva95, MardlingAarseth01, NaozFabrycky14}. However, there are no clear differences in the period distribution between the two samples. This is most surprising in the case of triple systems with short outer periods ($< 200$ days), raising the possibility that these systems are suspicious or have false orbital solutions. To explore this further, we reviewed the $41$ CHT candidates ($\Delta K > 5$) with outer periods shorter than $200$ days, finding that they are preferentially found around late-type stars. In contrast, the remaining triple candidates are typically found around early-type stars. Similarly, close binaries in the overall sample are also uncommon around cool stars. This pattern, assuming the orbital solutions are correct, suggests that for close triple candidates, the outer companion is likely a very low-mass star, which may help maintain the system’s dynamical stability \citep{MardlingAarseth01, NaozFabrycky14}.


The eccentricity distribution reveals that the compact triple candidates exhibit smaller eccentricities of the outer star $e_{\rm out}$ compared to all astrometric orbits. Systems with extremely eccentric outer orbits could experience potentially destabilising dynamical interactions with inner binaries, explaining the slightly `softer' eccentricity distribution.
In that case, the key measure of triple star stability is neither the orbital period ratio nor the eccentricity, but their combination \citep{MardlingAarseth01}.

\section{Close binary candidates in accelerated solutions}
\label{sec:accel}
 
In contrast to orbital solutions, where the astrometric orbit provides $K_0$ for comparison with our estimate of $K^*_1$, accelerated solutions lack this information. Therefore, the $\Delta K$ method used in the previous section to identify triple star candidates is not applicable here. Given that our goal is to distinguish between sources where the RV variations are caused by a distant accelerating companion and those resulting from an inner close binary within a triple system, we adopted an alternative approach to identifying potential close binary candidates among these accelerated solutions.

We adopted a methodology similar to that of \cite{Bashi24} and modelled the population of accelerated sources within the $G_{\mathrm{RVS}}-\log \left (\sigma_{\mathrm{RV}} \right )$ plane, where, $\sigma_{\mathrm{RV}}$ is the \gaia\  standard deviation of the epoch RV measurement defined in \citep{Katz23} as 
\begin{equation}
\sigma_{\mathrm{RV}} =\sqrt{\frac{2N_{\mathrm{RV}}}{\pi}\left(\zeta_{\mathrm{RV}}^2 - 0.11^2 \right)} ,
        \label{eq:sigmrv}
\end{equation}
where $\zeta_{\mathrm{RV}}$ is the uncertainty on the median of the
epoch RVs (\texttt{radial\_velocity\_error}) to which a constant shift of $0.11$~\kms~was added to take into account a calibration floor contribution, and $N_{\mathrm{RV}}$ is the number of transits used to derive the median RV (\texttt{rv\_nb\_transits}). 

Building upon the Bayesian approach described in Appendix~\ref{Appendix: A}, we modelled the accelerated star population using a density function composed of two Gaussian distributions: one representing the general accelerated solution sample ($\mathcal{N}^{\rm acc}_s$) and the other characterising close binary stars within the accelerated solutions ($\mathcal{N}^{\rm acc}_{b}$). The combined density function is defined as
\begin{equation}
\begin{aligned}
f(x | G_{\mathrm{RVS}}; \theta) &= (1 - F) \cdot \mathcal{N}^{\rm acc}_s\left(x | \mu_s(G_{\mathrm{RVS}}), \sigma^2_{\rm acc} (G_{\mathrm{RVS}})\right)\\
&+ F \cdot \mathcal{N}^{\rm acc}_b\left(x | \mu_b(G_{\mathrm{RVS}}), \sigma_b^2\right) ,
\end{aligned}
\label{eq:pdf_acc}
\end{equation}
where $x$ denotes the observed RV scatter $\log \left (\sigma_{\mathrm{RV}} \right)$. The parameter set $\theta$ includes the close binary fraction $F$, the parameters $a$, $b$, $G_{\rm min}$, $d$ that define the mean trends $\mu_s(G_{\mathrm{RVS}})$ and $\mu_b(G_{\mathrm{RVS}})$ (similar to the approach in Appendix~\ref{Appendix: A}; see Eqs.~\ref{eq:mus} and~\ref{eq:mub}), and the model’s expected standard deviations of the RV scatter for the accelerated stars and the close binaries within the accelerated sample, $\sigma_{\rm acc}(G_{\mathrm{RVS}})$ and $\sigma_b$, respectively.

A key distinction from the approach in the appendix is that we approximated the standard deviation, $\sigma_{\rm acc}$, of the accelerated sample by a linear function of $G_{\mathrm{RVS}}$, rather than assuming a constant value. This approximation accounts for the fact that closer (and brighter) sources with higher parallaxes exhibit larger angular separations for a given physical separation, making \gaia\  more sensitive to companions with smaller separations. Consequently, these systems tend to show larger RV variations due to accelerating companions. We defined
\begin{equation}
\log \sigma_{\mathrm{acc}}(G_{\mathrm{RVS}}) = m - n \, G_{\mathrm{RVS}}  ,
\label{eq:sigma_acc}
\end{equation}
where $n$ and $m$ are additional parameters introduced to capture this linear dependence.

We employed a Bayesian framework to model the full dependence, using priors listed in Table~\ref{tab:priors_acc}. By incorporating the $G_{\mathrm{RVS}}$-dependent RV scatter, our approach allows for a more conservative selection of close binary candidates, effectively accounting for the magnitude-dependent sensitivity of \gaia \ in detecting RV variability.
\begin{table}
        \centering
        \caption{Prior and posterior distributions of model parameters of close binary candidates in accelerated solutions.}
        \label{tab:priors_acc}
        \begin{tabular}{lcc} 
                \hline \hline
                Parameter & Prior & Posterior\\
                \hline
                $F$ & $\log \mathcal{U}(0.01,1)$ & $0.0825_{-0.0029}^{+0.0040}$ \\
                $a$ & $\mathcal{U}(-5,5)$&  $0.223_{-0.032}^{+0.022}$  \\
                $b$ & $\log \mathcal{U}(10^{-3},1)$ & $0.703_{-0.029}^{+0.016}$\\
                $G_{\mathrm{min}}$ & $\mathcal{U}(0,12)$ & $12.7_{-1.6}^{+1.7}$\\
        $d$ & $\log \mathcal{U}(10^{-4},100)$ & $12.82_{-0.62}^{+0.57}$ \\
        $n$ & $\log \mathcal{U}(0.01,1)$ & $0.0873_{-0.0013}^{+0.0025}$\\
                $m$ & $ \mathcal{U}(-10,10)$ & $0.262\pm 0.025$ \\
            $\sigma_b$ & $\log \mathcal{U}(0.01,10)$ & $0.4257_{-0.0061}^{+0.0085}$\\
                \hline
        \end{tabular}
\end{table}

We show in Fig.~\ref{fig:accel__fit} a scatter plot of the sample of accelerated sources on the $G_{\mathrm{RVS}}-\sigma_{\mathrm{RV}}$ plane with the dashed red line marking our best-fit model. The upper three dotted lines mark the 1, 2, and 3$\sigma$ uncertainties. We defined sources above the $3\sigma_{\mathrm{acc}}$ threshold (i.e. $\log \left (\sigma_{\mathrm{RV}} \right ) > {\left (\mu + 3 \sigma_{\mathrm{acc}} \right )}$) as probable close binary sources. Consequently, we are left with a total sample of $3,115$ close binary candidates in accelerated sources and $78,737$ other accelerated sources.
   \begin{figure}
   \centering
\includegraphics[width=0.48\textwidth] {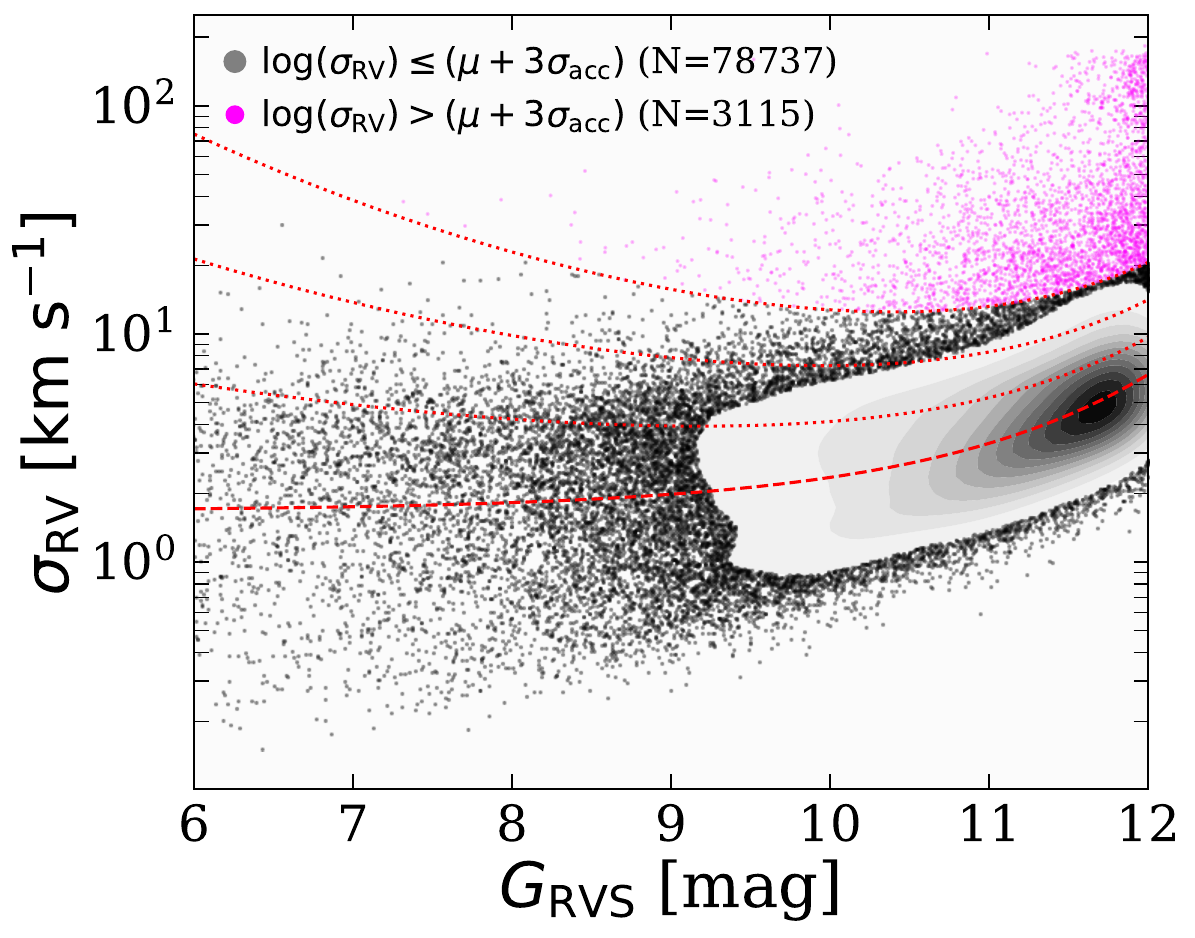}
   \caption{\gaia\  standard deviation of the epoch RV measurement as a function of RVS apparent magnitude $(\sigma_{\mathrm{RV}} - G_{\mathrm{RVS}})$ of accelerated \gaia\  sources. A dashed red line marks our best-fit curve using the Bayesian model of Eq.~\ref{eq:pdf_acc}. Dotted red curves mark the 1, 2, and $3\sigma_{\mathrm{acc}}$ uncertainties. Magenta points above $3\sigma_{\mathrm{acc}}$ are considered possible close-binary sources in accelerated solutions, while black points and respective contours indicate regions of higher data concentration of accelerated solutions.} 
              \label{fig:accel__fit}%
    \end{figure}

 \subsection{Candidate quality assessment}

Assuming the acceleration of the more luminous star in a binary system is given by
\begin{equation}
 \alpha_1 = \frac{GM_2}{a^2} ,
\end{equation}
where the semi-major axis $a$ is given by the Kepler's third law,
\begin{equation}
 a = \left (\frac{ G \left(M_1 + M_2 \right)P^2}{4 \pi ^2} \right ) ^ {1/3}\,\cdot
\end{equation}
We can then, using Eq.~\ref{eq:K_0} and the following relation 
\begin{equation}
 a_1=a\left(\frac{q}{1+q}\right),
\end{equation}
express the RV semi-amplitude $K_1$ of the primary star as
\begin{equation}
 K_1 =  G ^ {1/4} \alpha^{1/4}  \left (\frac{q}{1+q}\right)^{1/2} \frac{\sin i}{\sqrt{1-e^2}}\,\cdot
 \label{eq:K_alpha}
\end{equation}

In the case of accelerated \gaia  \ solutions, we only have information on the acceleration {projected} on the sky \citep{Halbwachs23}: 
\begin{equation}
 \Gamma =  \frac{\sqrt{g_{\alpha^*}^2 + g_{\delta}^2}} {\varpi} ,
\end{equation}
where $g_{\alpha^*}$ and $g_{\delta}$ are the accelerations in RA and Dec, respectively, and $\Gamma$ is expressed in units of AU $\mathrm{yr}^{-2}$.

Similar to our analysis of CHTs with astrometric orbits, we do not expect the typical RV peak-to-peak amplitude listed in the \gaia\  catalogue to fully reflect the amplitude of the outer orbit responsible for the acceleration. However, it can provide a rough estimate of velocity changes over a portion of the orbit, depending on the system’s configuration. Therefore, we can use Eq.~\ref{eq:K*_1} to approximate the velocity change, $s K^*_1$, where $s \leq 1$ is some positive value. Based on this and using Eq.~\ref{eq:K_alpha}, we can expect a dependence of the form $s K^*_1 \propto \Gamma^{1/4}$.


We show in Fig.~\ref{fig:acc_K1} a scatter plot of the accelerated sources (black points) and close binaries with a distant accelerated companion (magenta points). As this plot is displayed on a log scale, only positive $s K^*_1$ sources are considered. We fitted a two-dimensional Gaussian distribution to both subsamples and found a slope of $0.395$ for the accelerated sources, while there was almost no correlation with the triple candidates having a slope of $0.0064$. The lack of a clear correlation between $s K^*_1$ and $\Gamma$ in the high-$\sigma_{\rm RV}$ sample supports the conclusion that most of these sources with a few outliers with low $s K^*_1$ values are genuine triple-system candidates (i.e. close binaries within astrometric accelerations driven by an outer companion). Lastly, we note that while the observed slope of the accelerated solutions is larger than the theoretical expectation of $0.25$, this discrepancy likely arises from not accounting for complex selection effects in our sample, such as dependence on the distance (apparent magnitude). For nearby stars, \gaia \ detects smaller accelerations, leading to longer periods and smaller $K_1$. In fact, when we restricted our sample to fainter stars ($G_{\rm RVS} > 10$), the slope decreased to $0.31$.

   \begin{figure}
   \centering
\includegraphics[width=0.48\textwidth] {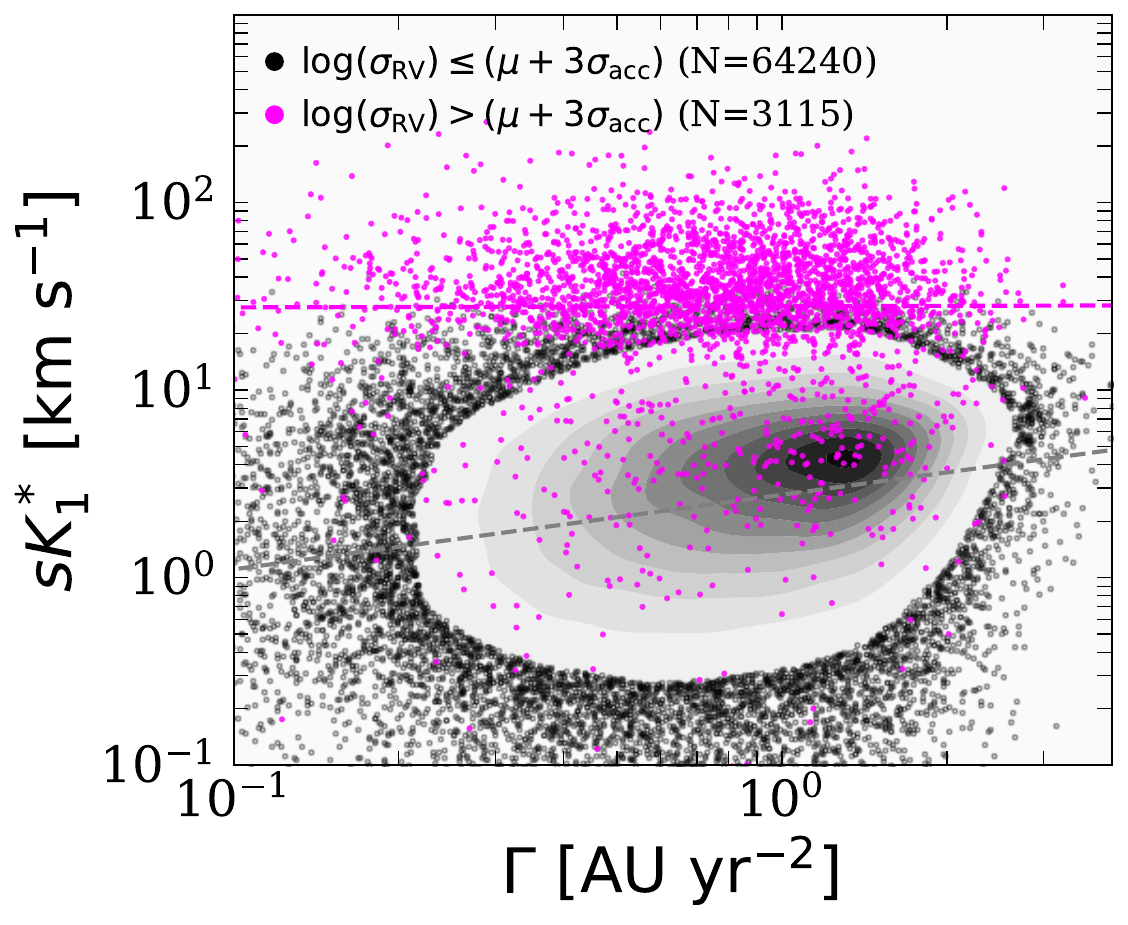}
   \caption{Semi-amplitude proxy as a function of the projected acceleration $(s K^*_1-\Gamma)$ of accelerated sources (black points and respective contours indicate regions of higher data concentration) and close binary candidates with a distant accelerated companion (magenta points). Dashed grey and magenta lines mark the slopes of the two samples using a two-dimensional Gaussian fit. }
              \label{fig:acc_K1}%
    \end{figure}

A partial list of all close binary candidates in accelerated solutions sorted by $\sigma_{\rm RV}$ is given in Table~\ref{tab:acc_rvsig_list}. The full table is available online.

\begin{table}
        \centering
        \caption{ Accelerated \gaia \  source IDs sorted by $\sigma_{\mathrm{RV}}$ (extract).}
 \small
        \label{tab:acc_rvsig_list}
        \begin{tabular}{cccc} 
                \hline \hline
                Gaia source id & $\Gamma$ & $s K^*_1$ & $\sigma_{\mathrm{RV}}$ \\

                  & [$\mathrm{AU yr}^{-2}$] & [\kms] & [\kms] \\
                \hline
1895909035311977344 & $0.79$ & $31.37$ & $183.02$ \\
5603843736049892864 & $0.74$ & $26.15$ & $174.02$ \\
1900881542288951552 & $1.17$ & $6.77$ & $173.99$ \\
1505576909891173760 & $0.59$ & $4.09$ & $173.33$ \\
3449974502473289728 & $0.79$ & $2.22$ & $171.03$ \\
5425165060658068352 & $2.41$ & $5.09$ & $170.13$ \\
2931992616121677056 & $2.10$ & $9.48$ & $169.54$ \\
4381157119149294592 & $1.08$ & $5.85$ & $169.32$ \\
1938074275445515008 & $0.39$ & $8.24$ & $167.87$ \\
203601172322941184 & $0.67$ & $11.44$ & $167.31$ \\
                \hline
\multicolumn{4}{l}{\footnotesize Note: The full table is available at the CDS.}
        \end{tabular}
\end{table}

 \subsection{Validation}
 \label{sec:accel_val}
 
While we are not able to validate the full triple star solutions, we can attempt to confirm the existence of an inner binary in some of the close binary candidates in the accelerated sample. As in the previous section, we followed three different approaches of validation with the MSC catalogue \citep{TokovininMSC97,TokovininMSC18} and cross-matching with binary catalogues \citep{GaiaEB23,Green23}.

\subsubsection{Cross-match with the Multiple Star Catalogue}

We performed a cross-match between the $3,115$ high-$\sigma_{\rm RV}$ sample and the $12,484$ systems listed in the MSC. This cross-match yielded a subsample of $7$ triple systems as well as $4$ quadruple systems.

\subsubsection{Eclipsing binaries catalogue}
We used the first \gaia\  catalogue of EB candidates \citep{GaiaEB23}, which lists a total of $2,184,477$ EB systems, and performed a cross-match based on \texttt{source\_id} with the sample of accelerated sources. We found that while there are only $18$ common sources out of a total of $78,737$ in the accelerated sample ($ \log \left (\sigma_{\mathrm{RV}} \right ) \leq \left( \mu + 3 \sigma_{\mathrm{acc}} \right )$), there are $198$ EB sources out of $3,115$ in the close binary candidates sample ($ \log \left( \sigma_{\mathrm{RV}} \right ) > {\left (\mu + 3 \sigma_{\mathrm{acc}} \right )}$). This represents an approximately $280$-fold higher fraction of EB systems in the close binary sample compared to the overall sample of accelerated sources. 

\subsubsection{Ellipsoidal catalogue}
We performed a cross-match with the ellipsoidal variable catalogue of \cite{Green23}. We find that while there are only $11$ common sources in the accelerated sample, this number increases to $55$ in the close binary candidate sample. The relative fraction between these two groups is approximately $130$ times higher in favour of binary candidates within the accelerated system sample.

\section{Discussion}
\label{sec:discussion}

In this work, we have compiled a list of $956$ CHT candidates found in \gaia\  astrometric binaries as well as $3,115$ close binary candidates found in astrometric accelerated \gaia\  solutions. To assemble this catalogue, we exploited available information on the peak-to-peak RV amplitude and variation of \gaia\  sources. We developed and utilised a metric, $\Delta K$, which compares the expected RV semi-amplitude variation with its astrometric semi-amplitude variation, adjusting for uncertainties. This approach effectively identifies CHT systems in which substantial RV variation arises predominantly from the inner binary rather than from the outer astrometric binary. In selecting close binary stars in accelerated systems, we further used information on the $\sigma_{\rm RV}$ of the system to assemble a highly significant subsample with large $\sigma_{\rm RV}$ among all accelerated \gaia\  solutions.

Our validation process involved cross-matching our candidates with known EBs and ellipsoidal variable catalogues, enhancing the reliability of our identifications. Additionally, the integration of RV data from ground-based surveys allowed us to set more precise orbital parameter constraints on these complex systems.



By exploring the CHT sample ($\Delta K > 5$), we find that the identified systems typically exhibit smaller eccentricities compared to all astrometric binaries, with a clear preference towards edge-on orbit orientations. The distribution of $\cos i$ provides evidence of some mutual orbit alignment in the CHT candidates. It is difficult to quantify the degree of such alignment without detailed modelling of all selection effects.

The findings of this study have several important implications. Firstly, the identified CHTs and triple candidates among accelerated sources contribute to a more comprehensive catalogue of multiple star systems, providing a valuable resource for further astronomical research and study. These systems offer a unique opportunity to study the formation and evolution of stars in dynamically complex environments.

Moreover, an important consideration in the analysis of such systems is the impact of blending effects, particularly in cases where the mass ratio, $q_{\rm out}$, exceeds $0.5$. Such configurations can lead to a reduction in the astrometric amplitude while leaving the RV amplitude seemingly unaffected, thereby introducing potential false positives in the identification of triple systems. This challenge is exacerbated when one star — typically the secondary — remains undetected in the RVS data, possibly due to rapid rotation or minor differences in magnitude. Addressing this issue requires further refinement to the detection and analysis techniques, possibly through the application of more sophisticated modelling approaches that can account for these subtleties.

Looking ahead, this work opens several avenues for further research. One immediate extension could be the exploration of additional multi-star configurations using catalogues of wide binaries \citep[e.g.][]{El-Badry21}. Such studies could uncover more intricate hierarchical structures and provide deeper insights into the stability and evolutionary pathways of multi-star systems. A large follow-up work is obviously needed to determine the inner periods and mass ratios of our sample of CHT candidates.

Moreover, the methodologies developed in this research could be adapted and applied to other datasets, potentially unveiling similar systems in different observational contexts. Future missions and surveys, which may offer higher precision and wider coverage, could significantly expand the number and types of detectable hierarchical systems.

In the longer term, the dynamics of these systems could also be computationally modelled to simulate their long-term evolution and potential outcomes. Such simulations could provide predictive insights into the behaviour of hierarchical systems and guide future observational strategies \citep[e.g.][]{MardlingAarseth01,NaozFabrycky14, Toonen20}.

\section*{Data availability}
The full versions of Tables~\ref{tab:DeltaK_list} and~\ref{tab:acc_rvsig_list} are available in electronic form at the CDS and at Zenodo using this link https://zenodo.org/records/14218818.

\begin{acknowledgements}
We thank the anonymous reviewers for their valuable comments and suggestions, which have significantly improved the quality of this paper.

We thank Ronny Blomme for a valuable discussion on the \gaia\  hot-star RV products.

D.B. acknowledges the support of the Blavatnik family and the British Friends of the Hebrew University (BFHU) as part of the Blavatnik Cambridge Fellowship and Didier Queloz for his worm hospitality.

This work has made use of data from the European Space Agency (ESA) mission {\it Gaia} (\url{https://www.cosmos.esa.int/gaia}), processed by the {\it Gaia}
Data Processing and Analysis Consortium (DPAC,
\url{https://www.cosmos.esa.int/web/gaia/dpac/consortium}). Funding for the DPAC
has been provided by national institutions, in particular the institutions
participating in the {\it Gaia} Multilateral Agreement.

Guoshoujing Telescope (the Large Sky Area Multi-Object Fiber Spectroscopic Telescope LAMOST) is a National Major Scientific Project built by the Chinese Academy of Sciences. Funding for the project has been provided by the National Development and Reform Commission. LAMOST is operated and managed by the National Astronomical Observatories, Chinese Academy of Sciences. 
This work used the Third Data Release of the GALAH Survey \citep{Buder21}. The GALAH Survey is based on data acquired through the Australian Astronomical Observatory, under programs:
A/2013B/13 (The GALAH pilot survey); A/2014A/25, A/2015A/19, A2017A/18 (The GALAH survey phase 1); A2018A/18 (Open clusters with HERMES); A2019A/1 (Hierarchical star formation in Ori OB1); A2019A/15 (The GALAH survey phase 2); A/2015B/19, A/2016A/22, A/2016B/10, A/2017B/16, A/2018B/15
(The HERMES-TESS program); and A/2015A/3, A/2015B/1,
A/2015B/19, A/2016A/22, A/2016B/12, A/2017A/14 (The HERMES K2-follow-up program). We acknowledge the traditional owners of the land on which the AAT stands, the Gamilaraay people, and
pay our respects to elders past and present. This paper includes data
that have been provided by AAO Data Central (datacentral.org.au).

Funding for the Sloan Digital Sky Survey IV has been provided by the Alfred P. Sloan Foundation, the U.S. Department of Energy Office of Science, and the Participating Institutions. SDSS acknowledges support and resources from the Center for High-Performance Computing at the University of Utah. The SDSS web site is www.sdss4.org.

SDSS is managed by the Astrophysical Research Consortium for the Participating Institutions of the SDSS Collaboration including the Brazilian Participation Group, the Carnegie Institution for Science, Carnegie Mellon University, Center for Astrophysics | Harvard \& Smithsonian (CfA), the Chilean Participation Group, the French Participation Group, Instituto de Astrofísica de Canarias, The Johns Hopkins University, Kavli Institute for the Physics and Mathematics of the Universe (IPMU) / University of Tokyo, the Korean Participation Group, Lawrence Berkeley National Laboratory, Leibniz Institut für Astrophysik Potsdam (AIP), Max-Planck-Institut für Astronomie (MPIA Heidelberg), Max-Planck-Institut für Astrophysik (MPA Garching), Max-Planck-Institut für Extraterrestrische Physik (MPE), National Astronomical Observatories of China, New Mexico State University, New York University, University of Notre Dame, Observatório Nacional / MCTI, The Ohio State University, Pennsylvania State University, Shanghai Astronomical Observatory, United Kingdom Participation Group, Universidad Nacional Autónoma de México, University of Arizona, University of Colorado Boulder, University of Oxford, University of Portsmouth, University of Utah, University of Virginia, University of Washington, University of Wisconsin, Vanderbilt University, and Yale University.

This research has made use of the SIMBAD database, CDS, Strasbourg Astronomical Observatory, France.

This research also made use of TOPCAT \citep{Taylor05}, an interactive
graphical viewer and editor for tabular data.

This work made use of Astropy,
a community-developed core Python package and an
ecosystem of tools and resources for astronomy \citep{Astropy22}.

\end{acknowledgements}

%
%
\bibliographystyle{aa}
\bibliography{aanda_acc} 

\begin{appendix} 
\section{Modelling \textit{Gaia}'s RV amplitude dependence on apparent magnitude}
\label{Appendix: A}

We followed a similar approach as \cite{Bashi24} and modelled the population of sources in the $G_{\mathrm{RVS}}$-$\log \left(\mathrm{RV_{pp}}/2 \right)$ plane. Our goal was to model the dependence of single stars, which do not show large RV  variations, to better characterise the impact of \textit{Gaia} RV errors on the source magnitude, independent of astrophysical effects.

We defined a density function as the sum of two Gaussian distributions: one for low-$\mathrm{RV_{pp}}/2$ (single stars) and one for high-$\mathrm{RV_{pp}}/2$ (binary stars). Each Gaussian was weighted by a binary fraction $F$. 
The mean of the single-star population was modelled using three free parameters, $a, b$, and  $G_0$, by the function 
\begin{equation}
\mu\left(x\right)  = a + e^{b(x-G_0)},
  \label{eq:mus}
\end{equation}
while the binary mean was modelled as
\begin{equation}
\mu_b(G_{\mathrm{RVS}}; a, ,b, G_{\mathrm{min}}, d) = \log_{10} {\left( \sqrt{\left(10^{\mu_s(G_{\mathrm{RVS}}; a, ,b, G_{\mathrm{min}})}\right)^2 + d^2}\right)},
        \label{eq:mub}
\end{equation}
where $d$ accounts for the extra RV variability in the binaries population.
The final density function is then
\begin{equation}
    \begin{aligned}
        f(x | G_{\mathrm{RVS}}; \theta) &= (1 - F) \cdot \mathcal{N}_s\left(x | \mu_s(G_{\mathrm{RVS}}), \sigma_s^2\right)\\
        &+ F \cdot \mathcal{N}_b\left(x | \mu_b(G_{\mathrm{RVS}}), \sigma_b^2\right),
    \end{aligned}
        \label{eq:pdf}
\end{equation}
where $\theta = (F, a, b, G_\text{min}, d, \sigma_s, \sigma_b)$ and $\sigma_s$ and $\sigma_b$ mark the standard deviations of the single and binary cases.

We then used a Bayesian framework similar to \cite{Bashi24} to model the full dependence using the priors listed on the left side of Table~\ref{tab:priors}. We selected \gaia\  bright RV sources within a distance of $1000$ pc and applied selection criteria similar to those used in our astrometric sample selection of Sect.~\ref{sec:Sample}. We then used the Python Markov chain Monte Carlo package \texttt{emcee} \citep{emcee} with $40$ walkers and $10^4$ steps to estimate the parameter values and their uncertainties that maximise the sample likelihood. 

We show in Fig.~\ref{fig:RVamp_Grvs} the best-fit curve to our sample using Eq.~\ref{eq:mus}. The posteriors are listed on the right side of Table~\ref{tab:priors}.

\begin{table}[htbp]
        \centering
        \caption{Prior and posterior distributions of model parameters of the binary fraction ($F)$, single and binary Gaussian means ($a,~b,~G_{\mathrm{min}},~d$), and stds ($\sigma_s,~\sigma_b$). }
        \label{tab:priors}
        \begin{tabular}{lcc} 
                \hline \hline
                Parameter & Prior & Posterior\\
                \hline
                $F$ & $\log \mathcal{U}(0.01,1)$ & $0.2295 \pm 0.0039$\\
                $a$ & $\mathcal{U}(-5,5)$& $-0.959\pm 0.017$ \\
                $b$ & $\log \mathcal{U}(10^{-3},1)$ & $0.2571_{-0.0039}^{+0.0033}$\\
                $G_{\mathrm{min}}$ & $\mathcal{U}(0,12)$ & $9.30\pm 0.67$ \\
        $d$ & $\log \mathcal{U}(10^{-4},100)$ & $9.31_{-0.19}^{+0.17}$\\
        $\sigma_s$ & $\log \mathcal{U}(10^{-3},10)$ & $0.1653_{-0.0008}^{+0.0010}$\\
            $\sigma_b$ & $\log \mathcal{U}(0.01,10)$ & $0.3902 \pm 0.0034$\\
                \hline
        \end{tabular}
\end{table}

   \begin{figure}[htbp]
   \centering
\includegraphics[width=0.5\textwidth] {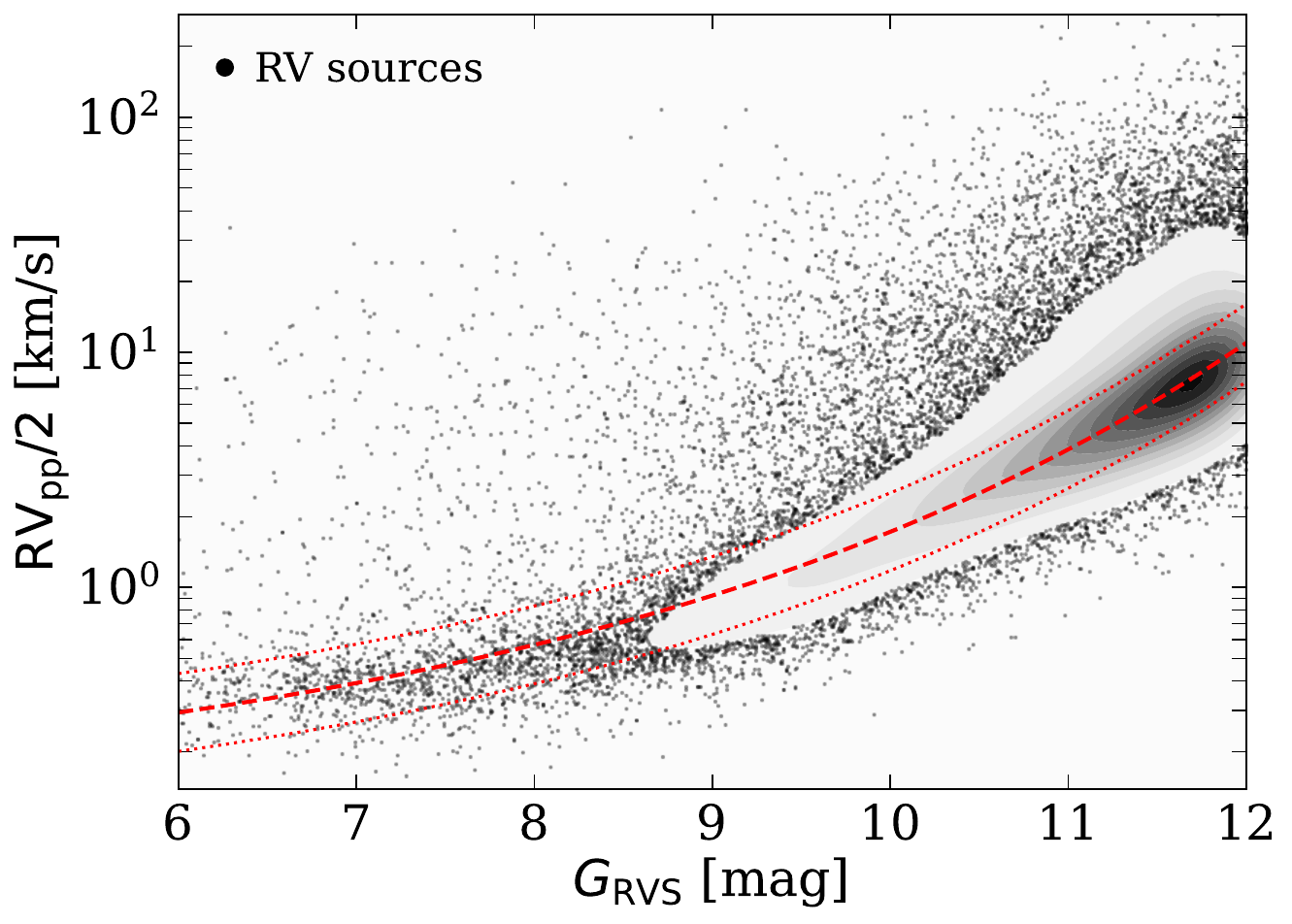}
   \caption{RV  variation as a function of the RVS apparent magnitude ($G_{\mathrm{RVS}}$-$\log \left(\mathrm{RV_{pp}}/2 \right)$) of bright \gaia\ RV sources. Respective contours indicate regions of higher data concentration. A clear monotonic trend between $G_{\mathrm{RVS}}$ and $\log \left(\mathrm{RV_{pp}}/2 \right)$ is
evident, with points well above this trend being suspected binary systems. The dashed red line marks our best-fit curve to our single star sample using Eq.~\ref{eq:mus}, while the two dotted curves mark the $\pm \sigma_s$ uncertainties.
}
    \label{fig:RVamp_Grvs}%
    \end{figure}

\clearpage

\section{Assessment of semi-amplitude estimation from RV variations}
\label{Appendix: B}

\subsection{Simulation}

To evaluate how well the peak-to-peak RV variation represents the actual semi-amplitude $K_1$, we simulated binaries, sampled their RV curves at uniformly distributed random phases, and computed the parameter $\mathrm{RV_{pp}}$. An eccentricity distribution $f(e) \sim \sin(e/\pi)$ was adopted.
As shown in Fig.~\ref{fig:RVPP/K1}, the robust RV half-amplitude never exceeds $K_1$ and, on average, under-estimates it, especially for a small number of RV measurements. The median ratio $\mathrm{ RV_{pp}}/(2 K_1)$ is 0.8 for 16 samples, typical for our data (the median number of RV measurements is 23, the median number of \texttt{rv\_visibility\_periods\_used} is 16). 

In our case, as \textit{Gaia } may not observe the entire orbital phase of an astrometric binary, especially for systems with longer periods ($P > 1000$ days), we might only capture part of the velocity curve, particularly if observations are taken near quadrature or away from the periastron in the case of eccentric orbits. Periods of inner binaries in astrometric systems are by default much less than 1000 days, while missing certain phases is accounted for by the simulation. So, this comment might refer only to acceleration candidates, where the simulation is not applicable anyway. 

   \begin{figure} [htbp]
   \centering
         \includegraphics[width=0.5\textwidth] {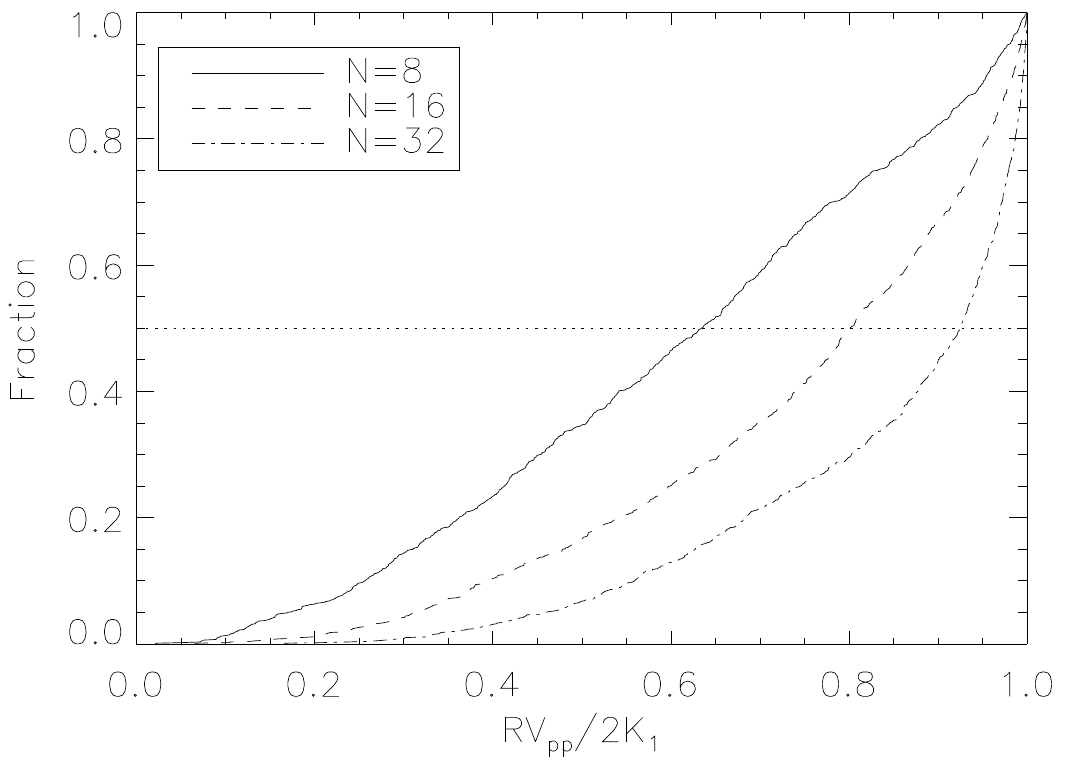}
        \caption{Cumulative distributions of the  $\mathrm{ RV_{pp}}/(2 K_1)$ ratio for simulated binaries sampled randomly 8, 16, and 32 times. }
              \label{fig:RVPP/K1}%
    \end{figure}

\subsection{Validation of $K^*_1$ estimates with the $S_{B^9}$ catalogue}

To assess the validity of our method and whether $K^*_1$ is a robust estimator of the true RV amplitude, we used the $S_{\mathrm{B^9}}$ catalogue \citep[][]{sb9} and compared the catalogue’s RV amplitude $K_1$ with our estimate of the semi-amplitude $K^*_1$. We selected a sample of binaries with orbital periods $P < 1200$ days and \texttt{Grade > 3} to exclude less reliable solutions. In total, we identified $184$ binary systems listed in the $S{\mathrm{B^9}}$ catalogue that also had an $\mathrm{RV_{pp}}$ value reported in \gaia. Using this sample, we present in Fig. \ref{fig:SB9_K1} a scatter plot of the $S_{\mathrm{B^9}}$ catalogue’s semi-amplitudes as a function of the estimated semi-amplitudes $K^*_1$ from Eq. \ref{eq:K*_1}, with the dashed red line marking the 1:1 ratio. Overall, given the uncertainties in the semi-amplitudes, we find good agreement between the values. This is expected given the short orbital periods and predominantly circular orbits of most sources in this sample, which provide good \gaia\  phase coverage for these systems. 

   \begin{figure}
   \centering
\includegraphics[width=0.5\textwidth] {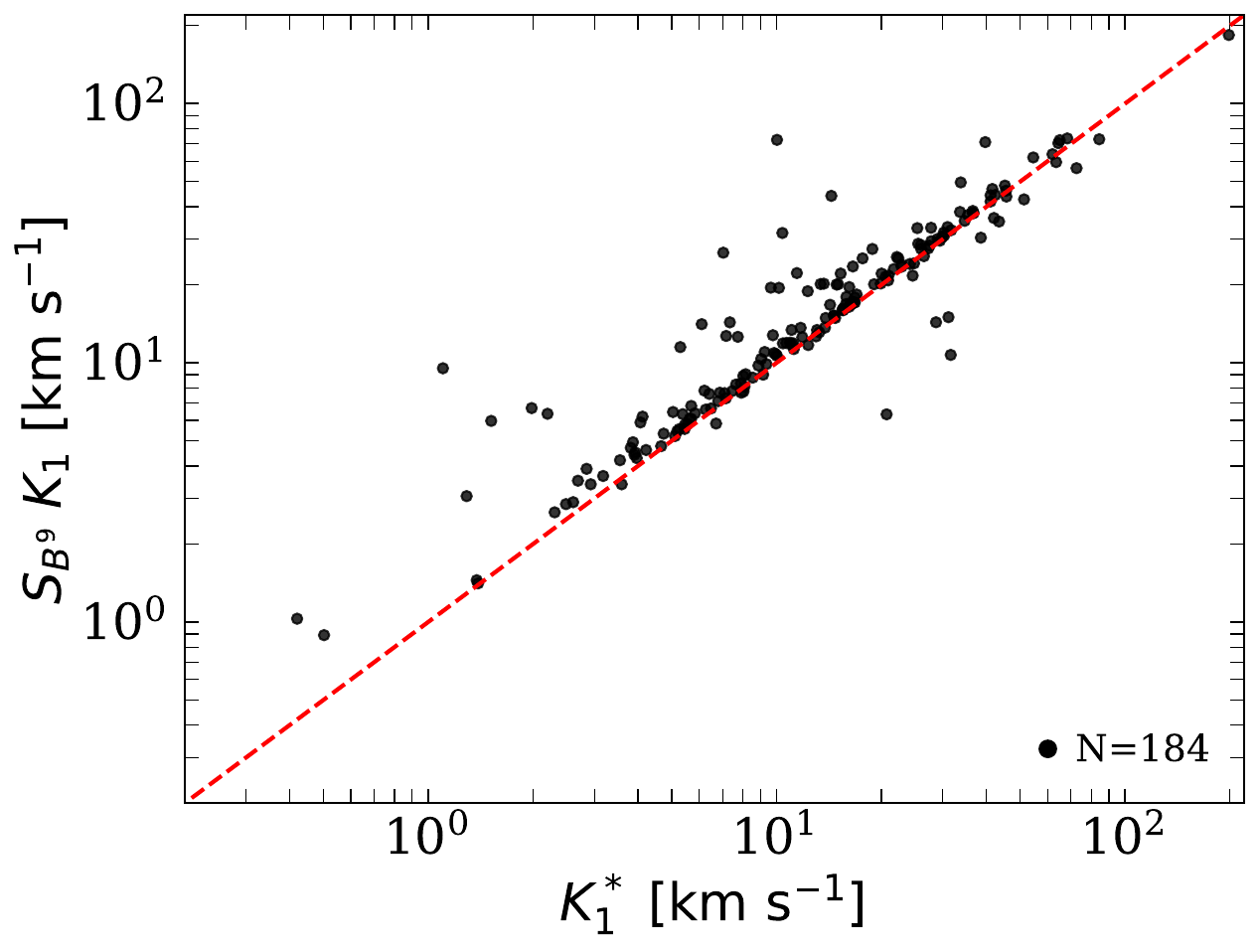}
   \caption{$S_{\mathrm{B^9}}$ catalogue's semi-amplitudes as a function of the estimated semi-amplitudes $(K^*_1 - S_{\mathrm{B^9}}~K_1)$ for binary stars in common. The calculation of $K^*_1$ is based on Eq.~\ref{eq:K*_1}. The dashed red line marks a 1:1 ratio.}
              \label{fig:SB9_K1}%
    \end{figure}
Overall, this demonstration supports our method of estimating binaries semi-amplitudes. When considering the uncertainties in $K_1$, we do not find many outliers. Our simulations in the previous section show that $K_1$ is rarely underestimated by large factors. Therefore, the points significantly above the red line merit investigation. Indeed, we find that two of the most extreme cases above the red line are double-lined spectroscopic binaries (SB2s). 

Conversely, the points below the line are physically implausible and indicate a problem in the estimation of $K^*_1$. We find four such sources that, after closer examination, are revealed to be Classical Cepheid Variable stars in binary system \citep{HerbigMoore52, Imbert96} as listed in \texttt{Simbad} \citep{Simbad00}: 
Gaia  DR3 2055014277739104896 (MW Cyg); Gaia\  DR3 2007201567928631296 (V* Z Lac); Gaia  DR3 470361114339849472 (V* RX Cam); and Gaia DR3 1820309639468685824 (* 10 Sge). 
While the RV variability observed by \gaia\  reflects mostly the Cepheid pulsations rather than orbital motion, this intrinsic variability affects the estimation of $K^*_1$. As an example, \gaia\  DR3 2055014277739104896 has $K^*_1= 20.7 \pm 1.7$\kms~while the reported $S_{B^9}$ value is $K_1 = 6.33$\kms. The reported $S_{B^9}$ variability is caused by the binary companion with an orbital period $P=439.41$ days while the RV variability observed by \gaia\  reflects mostly the Cepheid pulsations with $P=5.95$ day period and an amplitude of 19.2\kms according to Table~4 of \citet{Imbert96}.

\section{CHT candidates with ground-based RVs}
\label{Appendix: c}

\onecolumn
\begin{longtable}{c c c c c c c c c c }
\caption{\label{tab:groundRV} Triple star candidates with with multiple-epoch RVs from ground-based surveys.}\\
\hline\hline 
\gaia\  source id & $\widehat{\mathrm{RV}}$ & $K_0$ & $K^*_1$ & $\Delta K$ & Instrument & $N_{\mathrm{RV}}$ & $\Delta T$ & $\left\langle\mathrm{RV}\right\rangle$ & $\Delta \mathrm{RV}$ \\
                & [\kms]   & [\kms] & [\kms]  &           &            &                   & [day]      & [\kms]                               & [\kms]           \\
\hline
\endfirsthead
\caption{Continued.}\\
\hline\hline
\gaia\  source id & $\widehat{\mathrm{RV}}$ & $K_0$ & $K^*_1$ & $\Delta K$ & Instrument & $N_{\mathrm{RV}}$ & $\Delta T$ & $\left\langle\mathrm{RV}\right\rangle$ & $\Delta \mathrm{RV}$ \\
                & [\kms]   & [\kms] & [\kms]  &           &            &                   & [day]      & [\kms]                               & [\kms]           \\
\hline
\endhead
\hline
\endfoot
150798986120103808 & $-27.96$ & $8.14$ & $21.37$ & $5.99$ & LAMOST & $2$ & $690.11$ & $-41.82$ & $3.6$ \\
169307473369369600 & $1.03$ & $8.58$ & $54.51$ & $13.39$ & LAMOST & $2$  & $298.23$ & $10.29$ & $85.37$ \\
184164005769643392 & $-3.9$ & $3.18$ & $32.55$ & $9.2$ & LAMOST & $2$  & $979.29$ & $-1.91$ & $18.82$ \\
188284872267811456 & $79.84$ & $3.31$ & $11.85$ & $5.25$ & LAMOST & $4$  & $618.26$ & $83.51$ & $5.08$ \\
229469627204495744 & $-20.67$ & $2.33$ & $14.47$ & $6.48$ & LAMOST & $2$  & $679.13$ & $-35.52$ & $4.46$ \\
379619278688234624 & $-52.85$ & $7.1$ & $44.78$ & $12.46$ & LAMOST & $2$  & $1.03$ & $-59.89$ & $1.43$ \\
403959305033393920 & $-14.67$ & $7.04$ & $52.88$ & $13.39$ & LAMOST & $2$  & $47.89$ & $-20.24$ & $7.49$ \\
713081054945749248 & $-7.52$ & $5.39$ & $43.1$ & $12.22$ & LAMOST & $2$  & $655.25$ & $0.03$ & $12.04$ \\
724823117574564608 & $17.31$ & $7.28$ & $24.41$ & $6.37$ & LAMOST & $4$  & $1443.07$ & $20.12$ & $44.04$ \\
834297676421784704 & $-40.91$ & $11.26$ & $35.54$ & $8.01$ & LAMOST & $2$  & $743.01$ & $-22.71$ & $28.34$ \\
854376820330570112 & $-41.04$ & $4.36$ & $55.29$ & $17.07$ & LAMOST & $2$  & $585.34$ & $-50.83$ & $3.34$ \\
1001253088262279424 & $11.26$ & $5.71$ & $25.53$ & $6.4$ & LAMOST & $2$  & $462.73$ & $3.14$ & $4.5$ \\
1020251824555311104 & $18.76$ & $4.81$ & $19.8$ & $6.83$ & LAMOST & $5$  & $1403.16$ & $19.11$ & $10.27$ \\
1020642421765492480 & $-4.89$ & $5.67$ & $39.83$ & $12.8$ & LAMOST & $2$  & $693.16$ & $18.48$ & $7.39$ \\
1209993996408615680 & $39.57$ & $5.51$ & $25.58$ & $9.42$ & LAMOST & $2$  & $853.65$ & $35.94$ & $0.21$ \\
1275790936877524608 & $-46.28$ & $11.64$ & $76.6$ & $22.07$ & LAMOST & $2$  & $714.03$ & $-71.18$ & $131.53$ \\
1284938804897625600 & $4.96$ & $3.98$ & $16.89$ & $5.23$ & LAMOST & $2$  & $307.14$ & $-8.87$ & $8.07$ \\
1458572925243347584 & $-32.05$ & $10.76$ & $145.88$ & $19.22$ & LAMOST & $4$  & $1854.95$ & $10.65$ & $172.43$ \\
1466122133423051136 & $-11.41$ & $8.3$ & $28.31$ & $9.64$ & LAMOST & $4$  & $1814.0$ & $-29.79$ & $23.89$ \\
1476332718090832000 & $5.41$ & $7.39$ & $17.8$ & $5.3$ & LAMOST & $2$  & $60.87$ & $-1.82$ & $5.19$ \\
1489722742492254592 & $-48.68$ & $1.19$ & $51.79$ & $14.82$ & LAMOST & $2$  & $1472.99$ & $-26.15$ & $3.08$ \\
1493955656101094784 & $-30.21$ & $1.72$ & $20.32$ & $10.27$ & LAMOST & $2$  & $1020.21$ & $-35.33$ & $9.38$ \\
1499137207726399616 & $-35.47$ & $7.02$ & $28.13$ & $10.53$ & LAMOST & $2$  & $599.29$ & $-49.15$ & $4.8$ \\
1499334844941632512 & $-8.57$ & $1.84$ & $12.44$ & $5.43$ & LAMOST & $2$  & $1534.78$ & $-15.06$ & $5.08$ \\
1521516937979447040 & $-14.65$ & $3.35$ & $14.76$ & $5.87$ & LAMOST & $3$  & $104.7$ & $-15.83$ & $16.59$ \\
1560143713473045248 & $35.4$ & $13.07$ & $27.1$ & $5.95$ & LAMOST & $3$  & $1846.05$ & $24.83$ & $20.48$ \\
1574492580732435072 & $0.82$ & $6.6$ & $32.24$ & $8.97$ & LAMOST & $5$  & $1514.88$ & $4.48$ & $59.6$ \\
3067884732831947392 & $36.71$ & $4.86$ & $13.55$ & $5.86$ & LAMOST & $2$  & $332.09$ & $25.46$ & $14.48$ \\
3070571041597165440 & $-4.64$ & $3.86$ & $14.79$ & $5.09$ & LAMOST & $2$  & $333.06$ & $-4.12$ & $7.4$ \\
3095553049590375296 & $43.03$ & $7.05$ & $31.83$ & $8.84$ & LAMOST & $2$  & $655.17$ & $44.81$ & $20.7$ \\
3146157144545172096 & $3.02$ & $4.82$ & $26.34$ & $8.93$ & LAMOST & $2$  & $1130.85$ & $0.99$ & $29.14$ \\
3242212851168402816 & $3.56$ & $6.81$ & $40.64$ & $7.43$ & LAMOST & $2$  & $0.08$ & $1.62$ & $6.02$ \\
3269111097470951808 & $19.91$ & $4.43$ & $143.94$ & $11.04$ & LAMOST & $2$  & $0.03$ & $-0.75$ & $15.19$ \\
3293393395158188416 & $-25.81$ & $7.23$ & $36.54$ & $11.45$ & LAMOST & $2$  & $0.03$ & $-20.74$ & $1.82$ \\
3311594577501158272 & $0.85$ & $7.27$ & $49.95$ & $7.85$ & LAMOST & $2$  & $835.67$ & $9.33$ & $0.64$ \\
3340666558294094720 & $11.11$ & $5.3$ & $18.22$ & $5.72$ & LAMOST & $2$  & $390.01$ & $1.6$ & $0.06$ \\
145545725720154240 & $34.01$ & $6.66$ & $24.78$ & $7.74$ & GALAH & $2$  & $152.62$ & $30.17$ & $7.67$ \\
4656643472638236416 & $20.96$ & $11.6$ & $73.1$ & $14.41$ & GALAH & $2$  & $32.89$ & $28.55$ & $15.18$ \\
5268735063171919360 & $106.79$ & $6.44$ & $20.7$ & $5.6$ & GALAH & $2$  & $350.09$ & $104.56$ & $4.46$ \\
6243156524372883456 & $28.43$ & $9.44$ & $27.37$ & $7.6$ & GALAH & $2$  & $362.01$ & $15.55$ & $25.76$ \\
40119431248460032 & $44.06$ & $3.49$ & $83.58$ & $15.54$ & APOGEE & $4$  & $827.8$ & $23.02$ & $42.4$ \\
149025439503612800 & $35.53$ & $6.02$ & $22.45$ & $8.19$ & APOGEE & $4$  & $1797.96$ & $31.43$ & $50.25$ \\
225788290475544448 & $-24.23$ & $5.33$ & $20.43$ & $6.33$ & APOGEE & $2$  & $0.96$ & $-44.69$ & $0.51$ \\
834297676421784704 & $-40.91$ & $11.26$ & $35.54$ & $8.01$ & APOGEE & $3$  & $14.02$ & $-1.2$ & $16.59$ \\
842444370389914752 & $27.3$ & $6.4$ & $191.96$ & $18.99$ & APOGEE & $2$  & $7.01$ & $57.95$ & $7.78$ \\
878256460538469760 & $51.35$ & $5.27$ & $35.34$ & $8.13$ & APOGEE & $7$  & $9.98$ & $0.34$ & $60.85$ \\
1310339378926213120 & $2.2$ & $1.7$ & $13.77$ & $6.94$ & APOGEE & $7$  & $365.01$ & $15.51$ & $12.23$ \\
1314515289728827008 & $-28.28$ & $16.17$ & $78.0$ & $11.75$ & APOGEE & $2$  & $0.98$ & $12.02$ & $62.95$ \\
1318886943664522496 & $-16.46$ & $8.19$ & $53.63$ & $15.04$ & APOGEE & $3$  & $1.94$ & $3.02$ & $80.22$ \\
1327621915008445440 & $-73.85$ & $11.9$ & $27.11$ & $5.77$ & APOGEE & $5$  & $30.92$ & $-83.78$ & $14.6$ \\
1363264955245003136 & $-51.61$ & $8.52$ & $66.83$ & $12.6$ & APOGEE & $3$  & $305.17$ & $-53.95$ & $50.83$ \\
1437670212766194048 & $17.79$ & $10.97$ & $33.9$ & $7.12$ & APOGEE & $4$  & $99.7$ & $32.49$ & $64.24$ \\
1466122133423051136 & $-11.41$ & $8.3$ & $28.31$ & $9.64$ & APOGEE & $6$  & $1062.11$ & $-25.85$ & $73.37$ \\
1476332718090832000 & $5.41$ & $7.39$ & $17.8$ & $5.3$ & APOGEE & $6$  & $444.8$ & $15.3$ & $43.2$ \\
1489722742492254592 & $-48.68$ & $1.19$ & $51.79$ & $14.82$ & APOGEE & $6$  & $55.82$ & $0.22$ & $79.37$ \\
1492247770945997824 & $-12.31$ & $12.37$ & $26.46$ & $7.22$ & APOGEE & $3$  & $1.87$ & $-30.84$ & $3.15$ \\
1505152704561769856 & $-58.33$ & $5.33$ & $21.88$ & $8.95$ & APOGEE & $4$  & $10.03$ & $-64.22$ & $1.85$ \\
1555372692002194560 & $-28.77$ & $9.17$ & $60.5$ & $11.68$ & APOGEE & $2$  & $0.99$ & $7.3$ & $52.31$ \\
1560143713473045248 & $35.4$ & $13.07$ & $27.1$ & $5.95$ & APOGEE & $3$  & $446.74$ & $48.4$ & $24.65$ \\
1580848857452938368 & $8.78$ & $4.28$ & $13.24$ & $5.32$ & APOGEE & $5$  & $265.24$ & $7.28$ & $31.06$ \\
1656535217820593024 & $-36.52$ & $6.55$ & $23.87$ & $8.55$ & APOGEE & $2$  & $84.74$ & $-20.36$ & $46.62$ \\
2254632572254347264 & $-22.24$ & $3.58$ & $24.54$ & $10.42$ & APOGEE & $3$  & $2.01$ & $-15.85$ & $9.47$ \\
2474863445625114880 & $-31.11$ & $8.11$ & $17.19$ & $5.37$ & APOGEE & $5$  & $90.7$ & $-57.39$ & $24.37$ \\
2537538116668592000 & $-43.69$ & $7.16$ & $19.76$ & $6.65$ & APOGEE & $6$  & $302.12$ & $-18.98$ & $18.63$ \\
2779822783818066304 & $-8.71$ & $7.22$ & $51.35$ & $13.17$ & APOGEE & $5$  & $9.0$ & $10.04$ & $107.95$ \\
3016105118207579392 & $12.44$ & $4.84$ & $76.86$ & $14.06$ & APOGEE & $22$  & $2208.97$ & $38.51$ & $58.13$ \\
3230555004256693376 & $-13.52$ & $2.84$ & $13.71$ & $7.2$ & APOGEE & $3$  & $57.85$ & $9.3$ & $24.33$ \\
3311866122513508736 & $36.59$ & $5.48$ & $22.93$ & $6.7$ & APOGEE & $2$  & $1167.81$ & $48.86$ & $21.05$ \\
4651488030120521984 & $18.4$ & $7.59$ & $16.37$ & $5.02$ & APOGEE & $2$  & $363.01$ & $29.22$ & $10.87$ \\
4653901221916583936 & $16.24$ & $4.04$ & $17.24$ & $7.93$ & APOGEE & $10$  & $415.86$ & $8.38$ & $31.01$ \\
4655298456380769024 & $70.9$ & $5.97$ & $73.54$ & $12.22$ & APOGEE & $12$  & $1106.02$ & $22.29$ & $145.44$ \\
4705775115362757248 & $5.19$ & $4.99$ & $29.82$ & $8.69$ & APOGEE & $12$  & $11.96$ & $34.4$ & $10.91$ \\
6617696364275509760 & $7.82$ & $2.38$ & $10.07$ & $5.32$ & APOGEE & $2$  & $19.95$ & $35.87$ & $4.72$ \\
\end{longtable}
\twocolumn

\end{appendix}
\end{document}